\documentclass[12pt]{article}
\usepackage{amsmath}
\usepackage{graphicx,psfrag,epsf}
\usepackage{enumerate}
\usepackage{natbib}
\usepackage{amsthm,amsmath,color}
\usepackage{amssymb}
\usepackage{latexsym}
\usepackage{float}
\usepackage{amsfonts}
\usepackage{bbm}
\usepackage{longtable}
\usepackage{booktabs}
\usepackage{multirow} 
\usepackage{wrapfig}

\newcommand{\blind}{1}

\def\ds{\displaystyle}
\def\R{\mathbb{R}}
\newcommand{\argmin}[1]{ \underset{#1}{\text{argmin}} }

\def\tr{\text{tr}}

\begin{document}

\def\spacingset#1{\renewcommand{\baselinestretch}%
{#1}\small\normalsize} \spacingset{1}

%%%%%%%%%%%%%%%%%%%%%%%%%%%%%%%%%%%%%%%%%%%%%%%%%%%%%%%%%%%%%%%%%%%%%%%%%%%%%%

\title{\bf Generalized fiducial factor: an alternative to the Bayes factor for forensic identification of source problems}

\if1\blind
{\author{
Jonathan P Williams$^{(1)}$, Danica M Ommen$^{(2)}$, Jan Hannig$^{(3)}$ \\[.15in]
North Carolina State University$^{(1)}$ \\
Iowa State University$^{(2)}$ \\
University of North Carolina at Chapel Hill$^{(3)}$ \\
National Institute of Standards and Technology$^{(3)}$ \\
}} \fi
\maketitle

\vspace{-.15in}
\begin{abstract}
\small
One formulation of forensic identification of source problems is to determine the source of trace evidence, for instance, glass fragments found on a suspect for a crime.  The current state of the science is to compute a Bayes factor (BF) comparing the marginal distribution of measurements of trace evidence under two competing propositions for whether or not the unknown source evidence originated from a specific source.  The obvious problem with such an approach is the ability to tailor the prior distributions (placed on the features/parameters of the statistical model for the measurements of trace evidence) in favor of the defense or prosecution, which is further complicated by the fact that the typical number of measurements of trace evidence is typically sufficiently small that prior choice/specification has a strong influence on the value of the BF.  To remedy this problem of prior specification and choice, we develop an alternative to the BF, within the framework of generalized fiducial inference (GFI), that we term a {\em generalized fiducial factor} (GFF).  Furthermore, we demonstrate empirically, on the synthetic and real Netherlands Forensic Institute (NFI) casework data, deficiencies in the BF and classical/frequentist likelihood ratio (LR) approaches.
\end{abstract}

\vfill

\newpage
\spacingset{1.4}

\section{Introduction}

The adversarial nature of the criminal courtroom is extraordinarily troublesome in the context of Bayesian prior specification and choice.  In its purest form, subjectivist Bayesian theory \citep{savage1961,lindley1972} only admits prior probability distributions that reflect genuine beliefs about unknown features of a posited statistical model.  However, in the criminal courtroom setting there are inherently a range of such prior probability distributions that are reasonable, depending on the experts' role in the courtroom.  On one extreme there is the model representing the belief of the prosecution, and on the other extreme is the model representing the belief of the defense.  Further, given the high stakes nature of the outcome of a criminal court proceeding it is not hard to imagine that the subjectivist Bayesian inference from the evidence provided could lead to an extreme answer favoring either the prosecution or the defense, depending on which prior distribution is assumed for the statistical model features/parameters.  

Historically, the alternative to subjectivist Bayesian theory is to consider a class of {\em objective} prior distributions.  The problem with this approach is how to define {\em objective} in this context, and how to determine if the {\em objective} prior tends to favor the prosecution or the defense.  The critical question focuses on whether Bayesian methodology is actually appropriate for the criminal courtroom setting involving beliefs of expert witnesses (i.e., not only appropriate for each individual juror).  As statisticians, we have a responsibility to assess whether the methodological assumptions are safe and reliable.  To this end, we investigate a particular class of problems commonly referred to as forensic identification of source problems, and we motivate our work with a real data set of glass fragments that was gathered from 10 years of casework by the NFI \citep{es2017}.

Several approaches for assigning value to forensic evidence have been explored, including the Two-Stage approach \citep{parker1966, evett1977}, LRs with Bayesian treatment of parameter uncertainty \citep{lindley1977, evett1986, aitken2004} or with maximum likelihood estimates (MLE) of parameters \citep{grove1980, ommen2017_dissertation}, as well as score-based approaches \citep{gonzalez-rodriquez2005, egli2006, gonzalez-rodriquez2006, neumann2007, bolck2009, hepler2012}.  The BF approach is the most commonly recommended among European countries \citep{enfsi2015, taroni2016, berger2016, biedermann2016}, while a non-Bayesian approach is often recommended in the US \citep{swofford2018, kafadar2018}.  Recently, all of these methods have been scrutinized due to their lack of attention to the handling of uncertainty \citep{morrison2016, lund2017}.  In this paper we contribute to the discussion regarding how to handle uncertainty when quantifying the value of evidence, and we focus on a similar question to the one proposed in \cite{lund2017}: ``What do you really know versus what are you claiming to know (using prior information)?''

The gist of the LR approaches is to compare the probability of observing the evidence under two competing explanations for how the evidence was generated.  The Two-Stage approach, as it is most commonly presented, relies on statistical significance testing to compare two pieces of evidence; first to determine whether the evidence can be considered a ``match," and then to compare to other sources to determine how many others might also ``match".  This approach is not directly comparable to the recommended LR approaches \citep{shafer1982}, and will likely come under scrutiny due to the movement away from significance testing for applications with ``high-stakes" decisions \citep{wasserstein2016}.  The score-based likelihood ratio (SLR) approaches evolved from difficulties with the LR approaches for high-dimensional pattern and impression evidence (such as fingerprints, footwear, firearms, and handwriting evidence).  These SLR approaches rely on extensive training datasets consisting of pairwise comparison scores between evidential objects, and these scores can be created in a variety of different ways \citep{hepler2012, neumann2020-book, neumann2020}.  Again, this approach is not directly comparable to the recommended LR approaches due to the focus on modeling pairwise comparison scores as opposed to the features of one single object \citep{neumann2020-book}.  Due to the expressed concerns with the Two-Stage and SLR approaches, we will not consider these in this article.

Our contributions are the following.  First and most fundamentally, we develop methodology for a new solution to forensic identification of source problems based on the GFI approach \citep{hannig2016}.  It has been shown in the literature that GFI is asymptotically valid in the sense of Bernstein von-Mises type theory (again, see \cite{hannig2016}).  Second, we illustrate empirically via simulating the real NFI casework data that the BF can yield remarkably different answers when the priors reflect the prosecution instead of the defense hypotheses and vice versa, and that the BF values may be poorly calibrated to reflect the strength of evidence that they convey.  Our empirical results demonstrate very transparently that the degree to which the BF varies often may be more than enough to change the narrative of presented forensic evidence in a courtroom to the extent that a jury decision could conceivably be contrived.  Furthermore, an alternative LR statistic for this application is numerically unstable and poorly calibrated to these data.

GFI is a prior-free approach to estimating a posterior distribution which reflects the uncertainty associated with unknown model parameters.  We use GFI to define and construct the first ever GFF, particularly for application to statistical inference for forensic identification of source problems.  Moreover, we demonstrate in a real NFI data simulation that the GFF, which does not rely on prior specification, is able to provide meaningful, consistent, and well calibrated inference.  We make our R code and documentation for implementing the GFF publicly available at \verb1https://jonathanpw.github.io/software.html1.  The GFF can loosely be interpreted by analogy to a BF for particular choices of objective, data-driven priors, but the approach is justified independently of such interpretation.  However, the GFI, and by extension the GFF, approach has principled foundational roots in statistical theory.  We provide a gentle introduction to GFI prior to our construction of the GFF.  

The organization of the paper is as follows.  Section \ref{application} precisely defines and describes the context of forensic identification of source problems.  The real data is described and references are provided in Section \ref{data}.  Section \ref{methods} introduces the central notions for GFI, provides a brief overview of the established theory, and proceeds by deriving the necessary components for the GFF in the context of forensic identification of source for glass fragment data.  Thereafter, the main empirical results of the paper are presented in Section \ref{empirical_results}.  Finally, concluding remarks are provided in closing, and an appendix accounts specific details for the BF and LR.  The R code, along with a bash workflow file for reproducing all of our results is available at \verb1https://jonathanpw.github.io/research.html1.

\section{Motivating application}\label{application}

\begin{wrapfigure}{r}{.6\textwidth}
\vspace{-.15in}
\centering
\includegraphics[scale=.3]{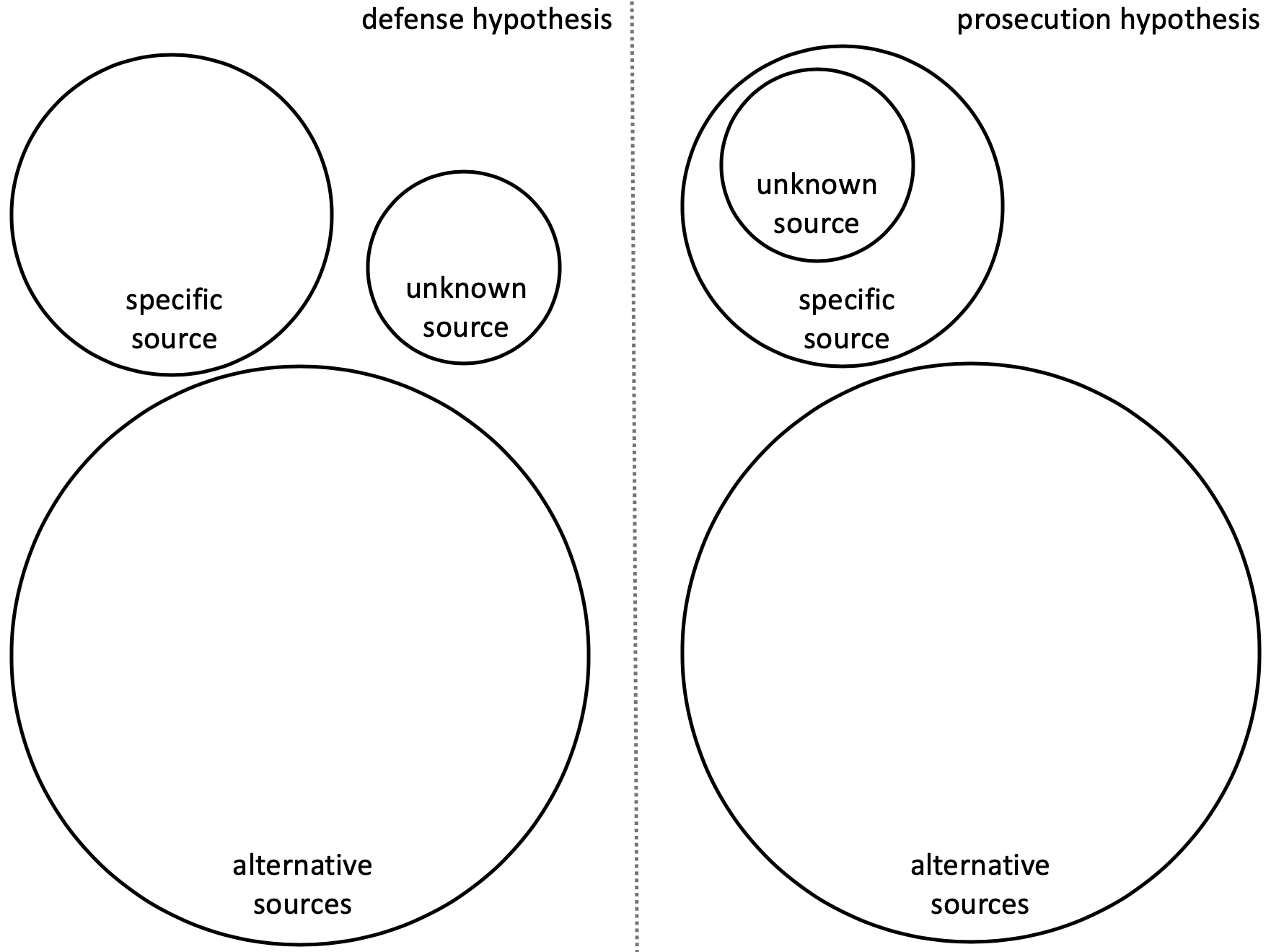}
\vspace{-.15in}\caption{\scriptsize Graphical description of the likely relationship between the specific, unknown, and alternative sources.}\label{sources_actual}
\vspace{-.15in}
\end{wrapfigure}

The motivation for the development of methodology for a GFF is the adversarial courtroom setting in which subjectivist BFs become problematic.  We focus our attention on the particular class of forensic identification of source problems.  The basic premise for such problems is that there is a crime that occurred at a specified location, and some evidential materials (e.g., blood, weapons, gunpowder, glass fragments, etc.) were found at the scene of the crime.  Next, a suspect for the crime is identified and is found with these same materials.  For example, glass fragments might be found at both the crime scene and fixed to the clothes of the suspect.  Perhaps the glass fragments are tiny, but nonetheless can be analyzed for chemical composition.  Then an important question involves assessing how likely it is that the glass fragments on the suspect originated from the window at the crime scene, which would link the suspect to the scene of the crime.

Within the context of forensic identification of source problems, we consider the following framework for constructing the competing hypotheses, sometimes referred to as the {\em specific source} formulation \citep{ommen2019}.  In this formulation, material evidence such as trace elements (i.e., the chemical composition of the chemical components that are useful for discrimination; see \cite{dettman2014}) of glass fragments found on a suspect are regarded as having been generated from either the {\em specific source} or some {\em alternative source}.  In the case of glass evidence, the data gathered from the suspect is regarded as having been generated from an {\em unknown source} (either the specific source at the crime scene or an alternative source often characterized by a background database), and the competing hypotheses are
\begin{quote}
\begin{description}
\item[$H_{p}$]: The unknown source evidence originated from the specific source.
\item[$H_{d}$]: The unknown source evidence originated from some other source in the alternative source population.
\end{description}
\end{quote}

We confine the rest of our exposition to modeling evidence arising from trace elements of glass fragments.  Figure \ref{sources_actual} gives a visual illustration of the likely relationship between the sources of data in the identification of source problem.  The alternative sources characterize a large database of panes of glass found in windows and doors used to describe the variation of trace element compositions found between and within panes of glass.  Glass fragments from a pane found at a specific source such as a crime scene also can be characterized based on the composition and variation of their trace elements.  When glass fragments are discovered on a suspect for a crime (i.e., the unknown source data), an analyst can compare the composition and variation of its trace elements to that of glass found at the specific source (i.e., the crime scene) and that of all types of glass that has been documented in the alternative source database.  This logical framework lends itself to modeling the alternative source data as a random effect, where the random effects component describes variation of trace elements between panes.  In Section \ref{methods} we formulate the construction of these data generating models.

Unfortunately, forensic databases are not sufficiently exhaustive for it to be realistic to assume that all relevant sources are represented in the alternative source data.  Nonetheless, the meaningful question for the forensic identification of source problem remains whether the unknown source data are consistent with the specific source data.  The alternative sources of data provide a benchmark for comparison.  In the sections that follow we develop and evaluate statistical methodology to address this question.  Further, we design a simulation study consistent with Figure \ref{sources_actual} from the real NFI casework data to investigate and assess our methods.

\section{Methodology}\label{methods}

The motivation for GFI is to construct prior-free probabilistic inference on meaningful parameters in a data generating model.  An overview of the ideas, common examples, and theoretical guarantees for GFI is presented in \cite{hannig2016}.  The formal definition of a GF distribution begins with a data generating equation $G$ for the realization of data $Y$, depending on some underlying pivotal quantity $U$ and some unknown fixed parameters $\theta$.  That is, $Y = G(U,\theta)$, where $G$ is deterministic and $U$ is a random variable whose distribution is completely known.  The idea for GFI is to invert the function $G$ to solve for the unknown parameters, and then switch the roles of $\theta$ and the observed data $y$ to construct a distributional estimator for $\theta$ that inherits the uncertainty associated with $U$.  

More precisely, consider the following inverse problem,
{\footnotesize
\[
Q_{y}(u^{\star}) = \argmin{\theta^{\star}}{\|y - G(u^\star, \theta^{\star})\|}.
\]
}For $\epsilon>0$, define the random variable $\theta_\epsilon^\star=Q_{y}(U_\epsilon^{\star})$,
where $U_\epsilon^{\star}$ has the same distribution as $U$ truncated to the set $\mathcal C_\epsilon = \{ U_\epsilon^{\star} : \|y - G(U_\epsilon^{\star}, \theta_\epsilon^\star )\| = \|y - G(U_\epsilon^{\star}, Q_{y}(U_\epsilon^{\star}) )\| \leq \epsilon \}$.  Then assuming that the random variables $\theta_\epsilon^\star$ converge in distribution as $\epsilon\to 0$, the GF distribution is defined as the limiting distribution $\theta^\star= \lim_{\epsilon\to 0} \theta_\epsilon^\star$. Notice that the fiducial distribution of $\theta^\star$ depends on the observed data $y$.  The intuition for understanding this distribution is similar to that for approximate Bayesian computations \citep{beaumont2002}.

Moreover, under certain conditions applicable to many practical settings \citep{hannig2016}, the GF distribution can be computed as
{\footnotesize
\begin{equation}\label{gfd}
r(\theta \mid y) = \frac{f(y \mid \theta) J(y,\theta)}{\int_{\Theta} f(y \mid \tilde{\theta}) J(y,\tilde{\theta}) \ d\tilde{\theta}}, 
\end{equation}
}where $f(y \mid \theta)$ is the likelihood function, and 
{\footnotesize
\begin{equation}\label{jacobian}
J(y,\theta) := D\bigg( \nabla_{\theta}G(u,\theta) \Big|_{u = G^{-1}(y,\theta)} \bigg),
\end{equation}
}with $D(A) := \sqrt{\det(A'A/n)}$, where $n$ is the number of samples observed (dimension of $y$).  The function $J(y,\theta)$ is a Jacobian-like quantity that results from inverting the data generating equation $y = G(U,\theta)$.  Viewed from another perspective, (\ref{gfd}) defines a posterior-like distribution for a class of data-driven, objective priors.  A variety of classes of objective (or non-informativie, weakly informative, etc.) priors are well-accepted in the literature and in practice as both meaningful and useful inferential tools \citep{jeffreys1946, bernardo1979, mukerjee1999, gelman2008, staicu2008, berger2009,martin2019}.  In fact, any prior distribution that is constructed for any reason other than to reflect the true state of the prior knowledge is not properly Bayesian.  In the following two subsections we use (\ref{gfd}) to construct GF distributions for the forensic identification of source problems described in the previous section.

\subsection{GF distribution of specific source data}\label{GF_s}

For the glass fragments found at the specific source, let $m$ denote the number of measurements of the log-transformed concentration of $p$ elements, and record the measurements as a column vector $y_{s,k} \in \R^{p}$ for $k \in \{1,\dots,m\}$.  Then, assuming a multivariate Gaussian data generating equation as in \cite{aitken2004} and \cite{ommen2017}, for $k \in \{1,\dots,m\}$,
{\footnotesize
\begin{equation}\label{dge_GF_s}
Y_{s,k} = G\big(Z_{k}, (\mu_{s}, A)\big) = \mu_{s} + AZ_{k},
\end{equation}
}where $Z_{k} \sim \text{N}_{p}(0,I_{p})$ and $A$ is nonsingular.  The GF distribution of $(\mu_{s}, A)$ then has the form,
{\footnotesize
\[
r_{s}\big(\mu_{s}, A \mid \{y_{s,k}\}\big) = \frac{  q_{s}\big(\mu_{s}, A \mid \{y_{s,k}\}\big)  }{  c_s  }, 
\]
}where $q_{s}\big(\mu_{s}, A \mid \{y_{s,k}\}\big) := f_{s}\big(\{y_{s,k}\} \mid \mu_{s}, A\big) \cdot J_{s}\big(\{y_{s,k}\}, (\mu_{s}, A)\big)$ is the unnormalized GF density with normalizing constant $c_{s}$, $f_{s}(\cdot\mid \mu_{s}, A) $ is a multivariate Gaussian density, and the Jacobian $J_{s}\big(\{y_{s,k}\}, (\mu_{s}, A)\big)$ is computed as follows.  As in \cite{Shi2017}, denoting $w := (y'_{s,1}, \dots, y'_{s,m})'$ and applying definition (\ref{jacobian}) gives $J_{s}\big(\{y_{s,k}\}, (\mu_{s}, A)\big)$, where
{\footnotesize
\[
\nabla_{(\mu_{s}, A)}G = 
\begin{pmatrix}
\frac{\partial w_{1}}{\partial (\mu_{s})_{1}} & \cdots & \frac{\partial w_{1}}{\partial (\mu_{s})_{p}} & \frac{\partial w_{1}}{\partial A_{11}} & \frac{\partial w_{1}}{\partial A_{12}} & \cdots & \frac{\partial w_{1}}{\partial A_{pp}} \\
\frac{\partial w_{2}}{\partial (\mu_{s})_{1}} & \cdots & \frac{\partial w_{2}}{\partial (\mu_{s})_{p}} &  \frac{\partial w_{2}}{\partial A_{11}} & \frac{\partial w_{2}}{\partial A_{12}} & \cdots & \frac{\partial w_{2}}{\partial A_{pp}} \\
\vdots & \ddots & \vdots & \vdots &\vdots &\ddots & \vdots \\
\frac{\partial w_{mp}}{\partial (\mu_{s})_{1}} & \cdots & \frac{\partial w_{mp}}{\partial (\mu_{s})_{p}} & \frac{\partial w_{mp}}{\partial A_{11}} & \frac{\partial w_{mp}}{\partial A_{12}} & \cdots & \frac{\partial w_{mp}}{\partial A_{pp}} \\
\end{pmatrix} = 
\begin{pmatrix}
I_{p} & I_{p}\otimes z'_{1} \\
\vdots & \vdots \\
I_{p} & I_{p}\otimes z'_{m} \\
\end{pmatrix}.
\]
}Rearranging rows of $\nabla_{(\mu_{s}, A)}G$ and denoting $\widetilde{U} := (z_{1},\dots,z_{m})'$ simplifies the expression to,
{\footnotesize
\[
\begin{split}
J_{s}\big(\{y_{s,k}\}, (\mu_{s}, A)\big) & = \bigg| 
\begin{pmatrix}
I_{p}\otimes 1'_{m} \\
I_{p}\otimes \widetilde{U}' \\
\end{pmatrix}
\begin{pmatrix}
I_{p}\otimes 1_{m} & I_{p}\otimes \widetilde{U} \\
\end{pmatrix}\bigg|^{\frac{1}{2}} m^{-\frac{p+p^{2}}{2}} \\
%& = \left|
%\begin{pmatrix}
%mI_{p} & I_{p}\otimes 1'_{m}\widetilde{U} \\
%I_{p}\otimes \widetilde{U}'1_{m} & I_{p}\otimes \widetilde{U}'\widetilde{U} \\
%\end{pmatrix} \right|^{\frac{1}{2}} \\
& = \left|
\begin{pmatrix}
I_{p} & 0 \\
0 & I_{p}\otimes A^{-1} \\
\end{pmatrix}
\begin{pmatrix}
mI_{p} & I_{p}\otimes 1'_{m}U \\
I_{p}\otimes U'1_{m} & I_{p}\otimes U'U \\
\end{pmatrix} 
\begin{pmatrix}
I_{p} & 0 \\
0 & I_{p}\otimes (A^{-1})' \\
\end{pmatrix}\right|^{\frac{1}{2}} m^{-\frac{p+p^{2}}{2}}, \\
\end{split}
\]
}where $1_{m}$ is an $m\times 1$ vector of ones, and $U := (y_{s,1} - \mu_{s}, \dots, y_{s,m} - \mu_{s})'$ so that $\widetilde{U} = U(A^{-1})'$.  Thus,
{\footnotesize
\[
q_{s}(\mu_{s}, A \mid \{y_{s,k}\}) = (2\pi)^{-\frac{mp}{2}} |AA'|^{-\frac{m+p}{2}}e^{-\frac{1}{2}\tr\big(S_{s}(AA')^{-1}\big)} \left|
\begin{pmatrix}
mI_{p} & I_{p}\otimes 1'_{m}U \\
I_{p}\otimes U'1_{m} & I_{p}\otimes U'U \\
\end{pmatrix}\right|^{\frac{1}{2}} m^{-\frac{p+p^{2}}{2}},
\]
}where 
{\footnotesize
\begin{equation}\label{S_s}
S_{s} := \sum_{k=1}^{m}(y_{s,k} - \mu_{s})(y_{s,k} - \mu_{s})'.
\end{equation}
}

\subsection{GF distribution of alternative source data}\label{GF_a}

For the glass fragments available in the alternative sources, let $m_{i}$ denote the number of measurements of the log-transformed concentration of $p$ elements for source $i \in \{1,\dots,n\}$, where $n$ is the total number of sources contained in the alternative source data.  Record the $p$ measurements as a column vector $y_{a,i,k} \in \R^{p}$ for $k \in \{1,\dots,m_{i}\}$ and $i \in \{1,\dots,n\}$.  Then, consistent with the specific source setup in the previous section, we assume that the data from each source in the alternative source data set is generated from a multivariate Gaussian distribution \citep{zadora2013} with a unique mean vector $\mu_{a} + Bt_{i}$, where $\mu_{a} \in \R^{p}$ is a fixed effect, and $Bt_{i} \in \R^{p}$ is a draw from a multivariate T random effect with $\tau$ degrees of freedom and positive-definite covariance matrix $BB'$ describing the variation in mean vectors over each source in the alternative source set.  The heavy tails of the multivariate T distribution reflect the inherently large variation that is observed in element composition exhibited by different panes of glass, while the light tails of the multivariate Gaussian distribution reflect the relatively small variance in element composition found in a single pane of glass.   

Accordingly, for $k \in \{1,\dots,m_{i}\}$ and $i \in \{1,\dots,n\}$,
{\footnotesize
\begin{equation}\label{dge_GF_a}
Y_{a,i,k} = \mu_{a} + BT_{i} + CV_{i,k},
\end{equation}
}where $V_{i,k} \sim \text{N}_{p}(0,I_{p})$, $C$ is nonsingular, and $T_{i} \sim \text{T}_{\tau}(0,I_{p})$.  Consequently, the GF distribution of $(\mu_{a}, B, C)$ can be expressed as,
{\footnotesize
\[
\begin{split}
r_{a}\big(\mu_{a}, B, C \mid \{y_{a,i,k}\}\big) & := \frac{q_{a}\big(\mu_{a}, B, C \mid \{y_{a,i,k}\}\big)}{c_{a}} \\
& = \frac{1}{c_{a}}\int\cdots\int q_{a}\big(\mu_{a}, B, C, \{t_{i}\}\mid \{y_{a,i,k}\}\big) \ dt_{1}\cdots dt_{n} \\
& = \frac{1}{c_{a}}\int\cdots\int q_{a}\big(\mu_{a}, B, C \mid \{t_{i}\}, \{y_{a,i,k}\}\big) f_{T_{1}}(t_{1})\cdots f_{T_{n}}(t_{n}) \ dt_{1}\cdots dt_{n} , \\
\end{split}
\]
}where $q_{a}\big(\mu_{a}, B, C \mid \{t_{i}\}, \{y_{a,i,k}\}\big) = f_{a}\big(\{y_{a,i,k}\} \mid \mu_{a}, B, C, \{t_{i}\}\big) \cdot J_{a}\big(\{y_{a,i,k}\}, (\mu_{a}, B, C)\big)$ is the unnormalized GF density with normalizing constant $c_{a}$, and $f_{a}\big( \cdot \mid \mu_{a}, B, C, \{t_{i}\}\big)$ is a multivariate Gaussian density.  To compute the Jacobian, as in the specific source derivation let $w := (y'_{a,1,1}, \dots, y'_{a,1,m_{1}}, \dots, y'_{a,n,1}, \dots, y'_{a,n,m_{n}})'$, denote $N := \sum_{i=1}^{n}m_{i}$, and apply definition (\ref{jacobian}) which gives $J_{a}\big(\{y_{a,i,k}\}, (\mu_{a}, B, C)\big)$, where
{\scriptsize
\[
\begin{split}
\nabla_{(\mu_{a}, B, C)}G %& = 
%\begin{pmatrix}
%\frac{\partial w_{1}}{\partial (\mu_{a})_{1}} & \cdots & \frac{\partial w_{1}}{\partial (\mu_{a})_{p}} & \frac{\partial w_{1}}{\partial B_{11}} & \frac{\partial w_{1}}{\partial B_{12}} & \cdots & \frac{\partial w_{1}}{\partial C_{11}} & \frac{\partial w_{1}}{\partial C_{12}} & \cdots & \frac{\partial w_{1}}{\partial C_{pp}} \\
%\frac{\partial w_{2}}{\partial (\mu_{a})_{1}} & \cdots & \frac{\partial w_{2}}{\partial (\mu_{a})_{p}} &  \frac{\partial w_{2}}{\partial B_{11}} & \frac{\partial w_{2}}{\partial B_{12}} & \cdots & \frac{\partial w_{2}}{\partial C_{11}} & \frac{\partial w_{2}}{\partial C_{12}} & \cdots & \frac{\partial w_{2}}{\partial C_{pp}}  \\
%\vdots & \ddots & \vdots & \vdots & \vdots & \ddots & \vdots & \vdots & \ddots & \vdots  \\
%\frac{\partial w_{N}}{\partial (\mu_{a})_{1}} & \cdots & \frac{\partial w_{N}}{\partial (\mu_{a})_{p}} & \frac{\partial w_{N}}{\partial B_{11}} & \frac{\partial w_{N}}{\partial B_{12}} & \cdots & \frac{\partial w_{N}}{\partial C_{11}} & \frac{\partial w_{N}}{\partial C_{12}} & \cdots & \frac{\partial w_{N}}{\partial C_{pp}}  \\
%\end{pmatrix} \\
& = 
\begin{pmatrix}
I_{p} & I_{p}\otimes t'_{1} & I_{p}\otimes v'_{1,1} \\
\vdots & \vdots  & \vdots \\
I_{p} & I_{p}\otimes t'_{1} & I_{p}\otimes v'_{1,m_{1}} \\
\vdots & \vdots  & \vdots \\
I_{p} & I_{p}\otimes t'_{n} & I_{p}\otimes v'_{n,1} \\
\vdots & \vdots  & \vdots \\
I_{p} & I_{p}\otimes t'_{n} & I_{p}\otimes v'_{n,m_{n}} \\
\end{pmatrix} \\
\end{split}
\]
}Next, rearranging rows of $\nabla_{(\mu_{a}, B, C)}G$ gives, 
{\footnotesize
\[
\begin{split}
J_{a}\big(\{y_{a,i,k}\}, (\mu_{a}, B, C)\big) & = \bigg| 
\begin{pmatrix}
I_{p}\otimes 1'_{N} \\
I_{p}\otimes W' \\
I_{p}\otimes \widetilde{Q}' \\
\end{pmatrix}
\begin{pmatrix}
I_{p}\otimes 1_{N} & I_{p}\otimes W & I_{p}\otimes \widetilde{Q} \\
\end{pmatrix}
\bigg|^{\frac{1}{2}} N^{-\frac{p+2p^{2}}{2}} \\
%& = \left|
%\begin{pmatrix}
%NI_{p} & I_{p}\otimes 1'_{N}W & I_{p}\otimes 1'_{N}\widetilde{Q} \\
%I_{p}\otimes W'1_{N} & I_{p}\otimes W'W & I_{p}\otimes W'\widetilde{Q} \\
%I_{p}\otimes \widetilde{Q}'1_{N} & I_{p}\otimes \widetilde{Q}'W & I_{p}\otimes \widetilde{Q}'\widetilde{Q} \\
%\end{pmatrix} \right|^{\frac{1}{2}} \\
%& = \left|
%\begin{pmatrix}
%I_{p} & 0 & 0 \\
%0 & I_{p^{2}} & 0 \\
%0 & 0 & I_{p}\otimes C^{-1} \\
%\end{pmatrix}
%\begin{pmatrix}
%NI_{p} & I_{p}\otimes 1'_{N}W & I_{p}\otimes 1'_{N}Q \\
%I_{p}\otimes W'1_{N} & I_{p}\otimes W'W & I_{p}\otimes W'Q \\
%I_{p}\otimes Q'1_{N} & I_{p}\otimes Q'W & I_{p}\otimes Q'Q \\
%\end{pmatrix} 
%\begin{pmatrix}
%I_{p} & 0 & 0 \\
%0 & I_{p^{2}} & 0 \\
%0 & 0 & I_{p}\otimes (C^{-1})' \\
%\end{pmatrix} \right|^{\frac{1}{2}} \\
& = \left|
\begin{pmatrix}
I_{p} & 0 & 0 \\
0 & I_{p^{2}} & 0 \\
0 & 0 & I_{p}\otimes (CC')^{-1} \\
\end{pmatrix}
\begin{pmatrix}
NI_{p} & I_{p}\otimes 1'_{N}W & I_{p}\otimes 1'_{N}Q \\
I_{p}\otimes W'1_{N} & I_{p}\otimes W'W & I_{p}\otimes W'Q \\
I_{p}\otimes Q'1_{N} & I_{p}\otimes Q'W & I_{p}\otimes Q'Q \\
\end{pmatrix} \right|^{\frac{1}{2}} N^{-\frac{p+2p^{2}}{2}}, \\
\end{split}
\]
}where {\scriptsize $W := 
\begin{pmatrix}
1_{m_{1}}\otimes t'_{1} \\
\vdots \\
1_{m_{n}}\otimes t'_{n} \\
\end{pmatrix}$}, and
{\scriptsize $
\widetilde{Q} := 
\begin{pmatrix}
v'_{1,1} \\
\vdots \\
v'_{1,m_{1}} \\
\vdots \\
v'_{n,1} \\
\vdots \\
v'_{n,m_{n}} \\ 
\end{pmatrix} = \underbrace{
\begin{pmatrix}
(y_{a,1,1} - \mu_{a} - B t_{1})' \\
\vdots \\
(y_{a,1,m_{1}} - \mu_{a} - B t_{1})' \\ 
\vdots \\
(y_{a,n,1} - \mu_{a} - B t_{n})' \\
\vdots \\
(y_{a,n,m_{n}} - \mu_{a} - B t_{n})' \\ 
\end{pmatrix}}_{\ds =: Q}(C^{-1})'.
$
}Thus, 
{\footnotesize
\[
\begin{split}
q_{a}(\mu_{a}, B, C \mid \{t_{i}\}, \{y_{a,i,k}\}) & = \frac{e^{-\frac{1}{2}\tr\big(S_{a}(CC')^{-1}\big)}}{(2\pi)^{\frac{pN}{2}} |CC'|^{\frac{N+p}{2}} N^{\frac{p+2p^{2}}{2}}} \left|
\begin{pmatrix}
NI_{p} & I_{p}\otimes 1'_{N}W & I_{p}\otimes 1'_{N}Q \\
I_{p}\otimes W'1_{N} & I_{p}\otimes W'W & I_{p}\otimes W'Q \\
I_{p}\otimes Q'1_{N} & I_{p}\otimes Q'W & I_{p}\otimes Q'Q \\
\end{pmatrix} \right|^{\frac{1}{2}} , \\
\end{split}
\]
}where 
{\footnotesize
\begin{equation}\label{S_a}
S_{a} := \sum_{i=1}^{n}\sum_{k=1}^{m_{i}} (y_{a,i,k} - \mu_{a} - B t_{i}) (y_{a,i,k} - \mu_{a} - B t_{i})'.
\end{equation}  
}

\subsection{Generalized fiducial factor}\label{GFF_section}

With the GF densities constructed for the specific source data in Section \ref{GF_s} and alternative source data in Section \ref{GF_a}, it remains to construct the GFF from them.  A key distinction between a BF and a GFF results from the fact that a prior distribution is necessarily independent of the data while the Jacobian, which is the analogue for the prior in GFI, is a function of the data.  To illustrate how this distinction breaks the construction of a BF, consider the data $y_{u,1},\dots,y_{u,m_{u}} \in \R^{p}$ from an unknown source, as described in Section \ref{application} (i.e., $m_{u}$ measurements of the log-transformed concentration of $p$ elements from glass fragments found on the suspect for a crime).  Let $M_{s}$ and $M_{a}$ denote the specific and alternative source models, respectively, and for conciseness let $\theta_{s} := (\mu_{s}, A)$ corresponding to the parameters for the specific source model (described in section \ref{GF_s}) and $\theta_{a} := (\mu_{a}, B, C)$ corresponding to the parameters for the alternative source model (described in section \ref{GF_a}).  Then the BF is defined as in \cite{kass1995} as,
{\footnotesize
\begin{equation}\label{BF}
\begin{split}
\text{BF} & := \frac{\pi\big(\{y_{u,j}\}\mid M_{s}\big)}{\pi\big(\{y_{u,j}\} \mid M_{a}\big)}
= \frac{\int \pi\big(\theta_{s}, \{y_{u,j}\} \mid M_{s}\big) \ d\theta_{s}}{\int \pi\big(\theta_{a}, \{y_{u,j}\} \mid M_{a}\big) \ d\theta_{a}} 
%& = \frac{\int f\big(\{y_{u,j}\} \mid \theta_{s}, M_{s}\big) \pi(\theta_{s} \mid M_{s}) \ d\theta_{s}}{\int f\big(\{y_{u,j}\} \mid \theta_{a}, M_{a}\big) \pi(\theta_{a} \mid M_{a}) \ d\theta_{a}} \\
= \frac{\int f_{s}\big(\{y_{u,j}\} \mid \theta_{s}\big)  \pi_{s}\big(\theta_{s} \mid \{y_{s,k}\}\big) \ d\theta_{s}}{\int f_{a}\big(\{y_{u,j}\} \mid \theta_{a}\big)  \pi_{a}\big(\theta_{a} \mid \{y_{a,i,k}\}\big) \ d\theta_{a}}. \\
\end{split}
\end{equation}
}This last equality does not make sense in the GFI paradigm in the same way that it would not make sense for an improper prior.  

The use of the conditional densities $\pi_{s}\big(\cdot \mid \{y_{s,k}\}\big)$ and $\pi_{a}\big(\cdot \mid \{y_{a,i,k}\}\big)$ requires them to be proper density functions (or at least integrable).  Nonetheless, the GF densities $r_{s}\big(\cdot \mid \{y_{s,k}\}\big)$ and $r_{a}\big(\cdot \mid \{y_{a,i,k}\}\big)$ are proper density functions, and share similar large-sample properties in the sense of Bernstein von-Mises type theory.  Hence, by analogy we define,
{\footnotesize
\begin{equation}\label{GFF}
\text{GFF} := \frac{\int f_{s}\big(\{y_{u,j}\} \mid \theta_{s}\big) \cdot r_{s}\big(\theta_{s} \mid \{y_{s,k}\}\big) \ d\theta_{s}}{\int f_{a}\big(\{y_{u,j}\} \mid \theta_{a}\big) \cdot r_{a}\big(\theta_{a} \mid \{y_{a,i,k}\}\big) \ d\theta_{a}},
\end{equation}
}and note the distinction from the BF.  In the remaining sections of this paper, we demonstrate empirically that the defined GFF has both practical utility for the identification of source problem and overcomes limitations of the BF and LR approaches.

\subsection{Remarks on computation}

In this section we describe our approach to compute the GFF defined in (\ref{GFF}) from actual data.  Applying the derivations of the GF distributions from Sections \ref{GF_s} and \ref{GF_a} directly into (\ref{GFF}) gives
{\footnotesize
\[
\text{GFF} = \frac{    \int\int f_{s}\big(\{y_{u,l}\} \mid \mu_{s}, A\big) \cdot r_{s}\big(\mu_{s}, A \mid \{y_{s,k}\}\big) \ d\mu_{s} \ dA  }{    E_{T_{1},\dots,T_{n+1}} \Big( \int\int\int f_{a}\big(\{y_{u,l}\} \mid \mu, B, C, t_{n+1}\big) \cdot \frac{1}{c_{a}}q_{a}\big(\mu_{a}, B, C \mid \{t_{i}\}, \{y_{a,i,k}\}\big) \ d\mu_{a} \ dB \ dC \Big)   }.
\]
}The numerator is the expected value of $f_{s}\big(\{y_{u,l}\} \mid \mu_{s}, A\big)$ (i.e., the specific source likelihood evaluated for the unknown source data) with respect to the GF density for the specific source model.  Accordingly, a natural estimate for this expected value is the average value of $f_{s}\big(\{y_{u,l}\} \mid \mu_{s}, A\big)$ over a GF sample of the parameters $\mu_{s}$ and $A$.  We thus construct a random walk Metropolis-Hastings Markov chain Monte Carlo (MCMC) algorithm to estimate a GF sample of $\mu_{s}$ and $A$.

The denominator is computationally much more difficult to deal with due to the expectation over the random effect components $T_{1},\dots,T_{n+1}$.  We have experimented with various strategies for importance sampling over all $T_{1}, \dots, T_{n+1}$, but these samples result in very poor mixing within the MCMC algorithm to estimate the GF distribution of $\mu_{a}$, $B$, and $C$.  A prohibitively large number of importances samples of the $\{T_{i}\}$ are needed to properly identify $BB'$ and $CC'$.  Accordingly, we construct the following point estimators for $t_{1}, \dots, t_{n}$.  

First, construct the estimates,  
{\footnotesize
\begin{equation}\label{estimates}
\begin{split}
\widehat{\mu}_{a} & := \frac{1}{N}\sum_{i=1}^{n}\sum_{k=1}^{m_{i}} y_{a,i,k} \\
\widehat{B}\widehat{B}' & := \frac{1}{n-1}\sum_{i=1}^{n} \big(\bar{y}_{a,i,\cdot} - \widehat{\mu}_{a}\big)\big(\bar{y}_{a,i,\cdot} - \widehat{\mu}_{a}\big)' \\
\widehat{C}\widehat{C}' & := \frac{1}{N-1}\sum_{i=1}^{n}\sum_{k=1}^{m_{i}}\big(y_{a,i,k} - \bar{y}_{a,i,\cdot}\big) \big(y_{a,i,k} - \bar{y}_{a,i,\cdot}\big)', \\
\end{split}
\end{equation}
}where $\bar{y}_{a,i,\cdot} := \frac{1}{m_{i}}\sum_{k=1}^{m_{i}}\bar{y}_{a,i,k}$ for each $i \in \{1,\dots,n\}$ and $\widehat{B}$ and $\widehat{C}$ are triangular Cholesky decomposition factors.  Then substituting these estimates into the data generating equation (\ref{dge_GF_a}) yields the following repeated observations regression model in terms of unknown coefficients $t_{i}$, $Y_{a,i,k} - \widehat{\mu}_{a} = \widehat{B}t_{i} + \widehat{C}V_{i,k}$, for $k \in \{1,\dots,m_{i}\}$ and $i \in \{1,\dots,n\}$.  Averaging over each measurement $k$ and rescaling the systems of equations gives the Gauss-Markov model,
{\footnotesize
\[
\big(\widehat{C}\widehat{C}'\big)^{-\frac{1}{2}}\big(\bar{Y}_{a,i,\cdot} - \widehat{\mu}_{a}\big) = \big(\widehat{C}\widehat{C}'\big)^{-\frac{1}{2}}\widehat{B}t_{i} + \big(\widehat{C}\widehat{C}'\big)^{-\frac{1}{2}}\widehat{C}\Bigg(\frac{1}{m_{i}}\sum_{k=1}^{^{m_{i}}}V_{i,k}\Bigg),
\]
}with the resulting least squares solution $\widehat{t}_{i} := \Big(\widehat{B}'\big(\widehat{C}\widehat{C}'\big)^{-1}\widehat{B}\Big)^{-1}\widehat{B}'\big(\widehat{C}\widehat{C}'\big)^{-1}\big(\bar{y}_{a,i,\cdot} - \widehat{\mu}_{a}\big)$, for every source $i \in \{1,\dots,n\}$ in the alternative source data set.  Note that $\widehat{t}_{i}$ is a consistent estimator for $t_{i}$ for large $m_{i}$.

Using the estimates $\{\widehat{t}_{i}\}$ we estimate the GFF as 
{\footnotesize
\[
\text{GFF} = \frac{    \int\int f_{s}\big(\{y_{u,l}\} \mid \mu_{s}, A\big) \cdot r_{s}\big(\mu_{s}, A \mid \{y_{s,k}\}\big) \ d\mu_{s} \ dA  }{   \int\int\int E_{T_{n+1}} \Big(f_{a}\big(\{y_{u,l}\} \mid \mu, B, C, t_{n+1}\big)\Big) \cdot \frac{1}{c_{a}}q_{a}\big(\mu_{a}, B, C \mid \{\widehat{t}_{i}\}, \{y_{a,i,k}\}\big) \ d\mu_{a} \ dB \ dC   }, \\
\]
}where the expectation $E_{T_{n+1}}(\cdot)$ is estimated by evaluating the average of the integrand over some large number of importance samples of $T_{n+1} \sim \text{T}_{5}(0,I_{p})$.  

The computation of the ratio of marginal densities such as a BF or the GFF is a difficult endeavor and a well explored topic in the literature \citep{meng1996,diciccio1997,gelman1998}.  Other popular approaches include importance, bridge, and path sampling \citep{gelman1998}, but we find that these methods nonetheless tend to require a fair amount of finesse and tailoring to a given data model.  The remaining sections of this paper serve to evaluate the empirical performance of our proposed GFF, and to illustrate shortcomings in the BF and LR.  The real data are described next

\subsection{NFI casework data}\label{data}

The glass fragment data set that we investigate \citep{es2017} was kindly received from the NFI, but the NFI was not further involved in this research.  Currently, these data are available on request by emailing p.zoon@nfi.nl.  

The data set consists of fragments from 979 unique windows from crime scenes spanning approximately 10 years of casework \citep{es2017}.  Of the 979 sources, 659 are designated as training data and the remaining 320 as calibration data.  Measurements of the log-transformed concentration of 18 elements for three fragments are recorded for the glass corresponding to each crime scene window in the training data, for a total of $3\times 659$ measurements.  Measurements of the 18 elements for five fragments for each window in the calibration data are recorded, for a total of $5\times 320$ measurements.  As discussed in \cite{es2017}, a meaningful subset of 10 of the 18 elements are considered.  Further details of these data are documented in \cite{es2017}.

In the context of our formulation (i.e., Figure \ref{sources_actual}), the training data corresponds to the alternative source data.  We then separate the first three measurements of each source in the calibration data set to denote a set of specific source data (each set corresponding to one unique window as the specific source), and leave the remaining two measurements to comprise sets of unknown source data.  Accordingly, we have 320 observed instances in which the unknown source is the specific source (i.e., the prosecution hypothesis, $H_{p}$), and $320\times 319$ observed instances in which the unknown source is {\em not} the specific source (i.e., the defense hypothesis, $H_{d}$).  We study our methods by simulating over these data and evaluating the performance of the GFF we construct, compared to the truth and compared to the BF and LR.

\section{Empirical results}\label{empirical_results}

In the empirical analysis that follows, we first demonstrate that all three methods (GFF, BF, and LR) perform well on fully synthetic data simulated from the data generation equations (\ref{dge_GF_s}) and (\ref{dge_GF_a}) when there are many specific and unknown source data measurements available.  Next, we illustrate the performance of all three methods in a similar simulation design but with only three data points in the specific source data sets, and two in the unknown source data sets.  This second simulation design allows us to exhibit the behavior of the GFF, BF, and LR using data generation equations (\ref{dge_GF_s}) and (\ref{dge_GF_a}), but with sample sizes the same as in the real NFI data.  Lastly, to assess performance using the real data we show the results of a simulation design that simply samples data sets from the real NFI data.  

Preprocessing of the data is described next, followed by a summary of each of the three simulation designs.  The results are presented and discussed in the subsections that remain.  The implementation of the BF follows as described in \cite{ommen2017} and \cite{ommen2019} (see their {\em specific source} formulation).  The LR is constructed from Chapter 7.2 of \cite{ommen2017_dissertation}.  For reference, the exact details are presented in the appendix.

\begin{table}[H]
\footnotesize
\begin{tabular}{l | llllllllll}
element & Ti49 & Sr88 & K39 & Zr90 & Mn55 & Ba137 & Ce140 & La139 & \textbf{Pb208} & \textbf{Rb85} \\
\hline
st dev & .00000 & .00001 & .00001 & .00001 & .00001 & .00002 & .00006 & .00007 & \textbf{.00012} & \textbf{.00013} \\
\end{tabular}
\caption{\scriptsize Sample standard deviation (rounded to five decimal places) of each element over all $3 \times 659$ measurements in the NFI training data set.  The data vector for each element was first rescaled to have unit Euclidean norm.}\label{elements}
\vspace{-.15in}
\end{table}

A limitation of the NFI casework data is that each specific source consists of only three measurements of the glass fragments making it difficult to obtain very reliable estimates of the specific source parameters, $\mu_{s}$ and $A$, regardless of the statistical framework (i.e., Bayesian, frequentist, or GF).  Since each of the three measurements records the log-transformed concentration of 10 elements (down from the original 18 as in \cite{es2017}), with so few measurements, this is in fact a relatively high-dimensional inference problem.  Moreover, since the unknown source data only consists of two measurements, consistent with a sure independence screening strategy \citep{fan2008}, in our analysis we reduce the dimension of the measurements to reflect only the two elements (i.e., $p = 2$) that exhibit the largest variation (after being rescaled to have unit norm) over all sources and measurements in the alternative source data set ($3 \times 659$ measurements in total).  Table \ref{elements} presents the variance observed for the rescaled, log-transformed concentrations of each of the 10 elements, from which we select elements ``Pb208'' and ``Rb85''.

% Summary of synthetic ideal data size simulation design
In the first simulation design we generate $n = 659$ alternative sources of data from (\ref{dge_GF_a}) with $m_{i} = 3$ measurements for each source.  The values of $\mu_{a}$, $B$, and $C$ used to generate the data are computed from the real NFI alternative source data via the equations in (\ref{estimates}).  Next, 320 specific source data sets are generated from (\ref{dge_GF_s}), each with $m = 150$ measurements.  Each of the 320 specific source data sets are generated from unique values of $\mu_{s}$ and $A$, each corresponding to a particular source of the 320 specific sources in the real NFI data set and computed as
{\footnotesize
\[
\begin{split}
\widehat{\mu}_{s} & := \frac{1}{m}\sum_{k=1}^{m} y_{s,k} \\
\widehat{A}\widehat{A}' & := \frac{1}{m-1}\sum_{k=1}^{m}\big(y_{s,k} - \widehat{\mu}_{s}\big) \big(y_{s,k} - \widehat{\mu}_{s}\big)'. \\
\end{split}
\]
}

To simulate $H_{p}$ true and $H_{d}$ true events, respectively, we must generate additional data with unknown sources.  For $H_{p}$ true, an additional $m_{u} = 2$ measurements for each of the 320 specific sources are generated from (\ref{dge_GF_s}) using the respective, previously computed values of $\mu_{s}$ and $A$.  For $H_{d}$ true, an additional 3,000 sets of $m_{u} = 2$ measurements are generated the same as the alternative sources of data.  Accordingly, 320 simulated GFF, BF, and LR values for $H_{p}$ true are computed using the 320 pairs of unknown and specific source data sets, and 3,000 simulated GFF, BF, and LR values for $H_{d}$ true are computed using 3,000 non-associated pairs of unknown and specific source data sets (the specific sources are randomly selected from among the 320, for each of the 3,000 unknown sources).  

While we could have generated only one data set of $n = 659$ alternative sources of data and one set of 320 specific sources of data, to account for variation in these sources a new set is generated for each of the 3,320 simulated events.  The LR crashed for one of the 3,000 simulated $H_{d}$ true events, so for comparison sake, we omit the data associated with this random number generator seed for all three simulation designs (i.e., all simulation designs have data for 2,999 data sets for $H_{d}$ true).  We describe this simulation design as having ideal sample sizes because $m = 150$ whereas $m = 3$ for the real NFI data.  This difference has a particularly significant effect on the stability of the LR, as will be seen in the two simulation designs that follow.  See the results in Section \ref{simulation1}.

% Summary of synthetic actual data size simulation design 
This second simulation design is the same as the first, with the modification being that the specific sources each contain only $m = 3$ measurements, as is the case for the real NFI data set.  Thus, this simulation is designed to observe the performance of the GFF, BF, and LR on synthetic data that most closely resembles the real NFI data.  See the results in Section \ref{simulation2}.

% Summary of real data simulation design
The third simulation design uses the measurement values from real NFI data set.  Recall from Section \ref{data} that for each of the 320 specific sources (each containing $m = 3$ measurements) there are an associated two held out measurements.  With these 320 sets of $m_{u} = 2$ measurements each serving as unknown sources, we are able to simulate 320 $H_{p}$ true events and $320\times 319$ $H_{d}$ true events.  For comparison with the first and second simulation designs, however, we only sample a random subset of 3,000 of the $320\times 319$ $H_{d}$ true events.  See the results in Section \ref{simulation3}.

\subsection{Simulation 1: fully synthetic data with ideal sample sizes}\label{simulation1}

\begin{wrapfigure}{r}{0.75\textwidth}
\vspace{-.15in}
\centering
\includegraphics[scale=.35]{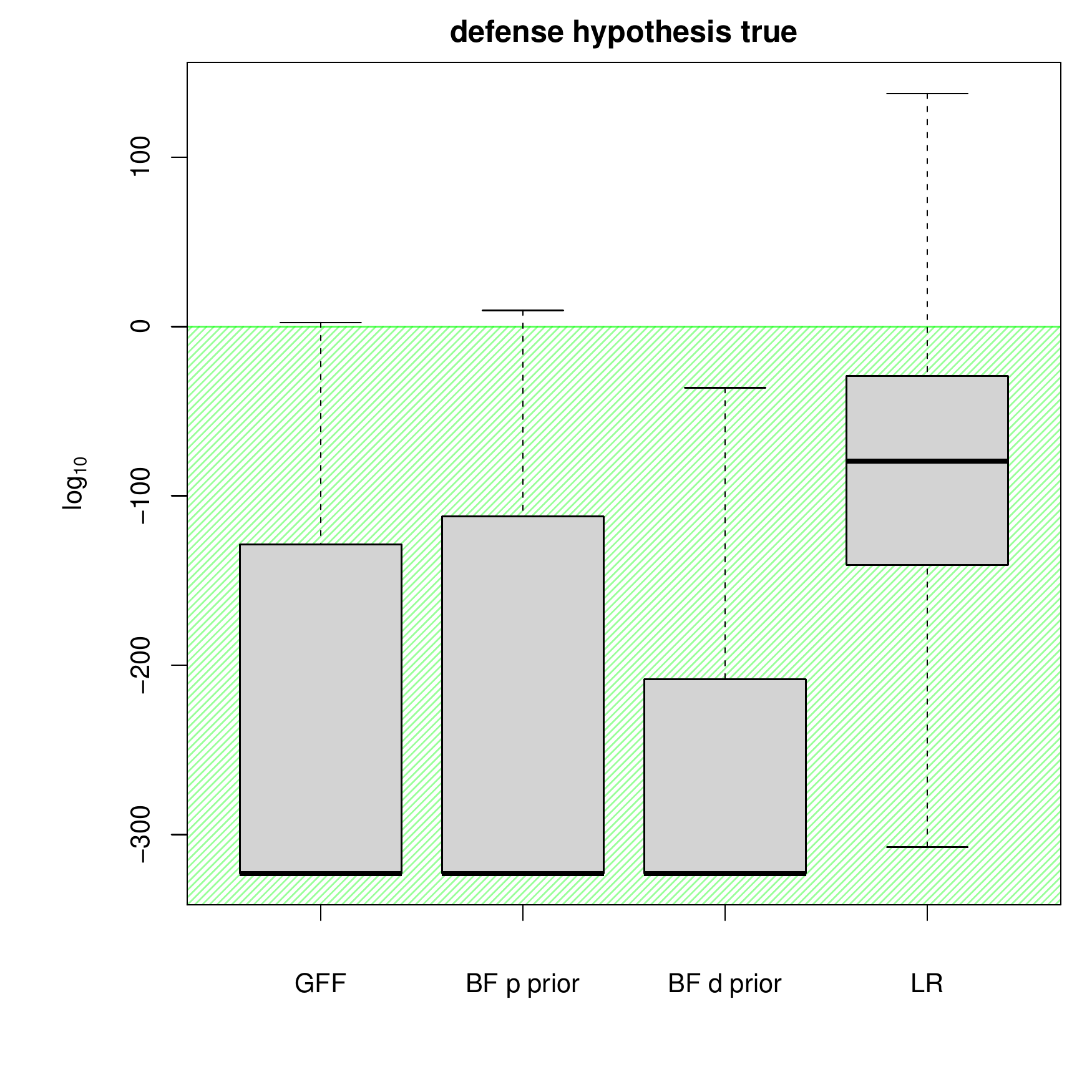}\includegraphics[scale=.35]{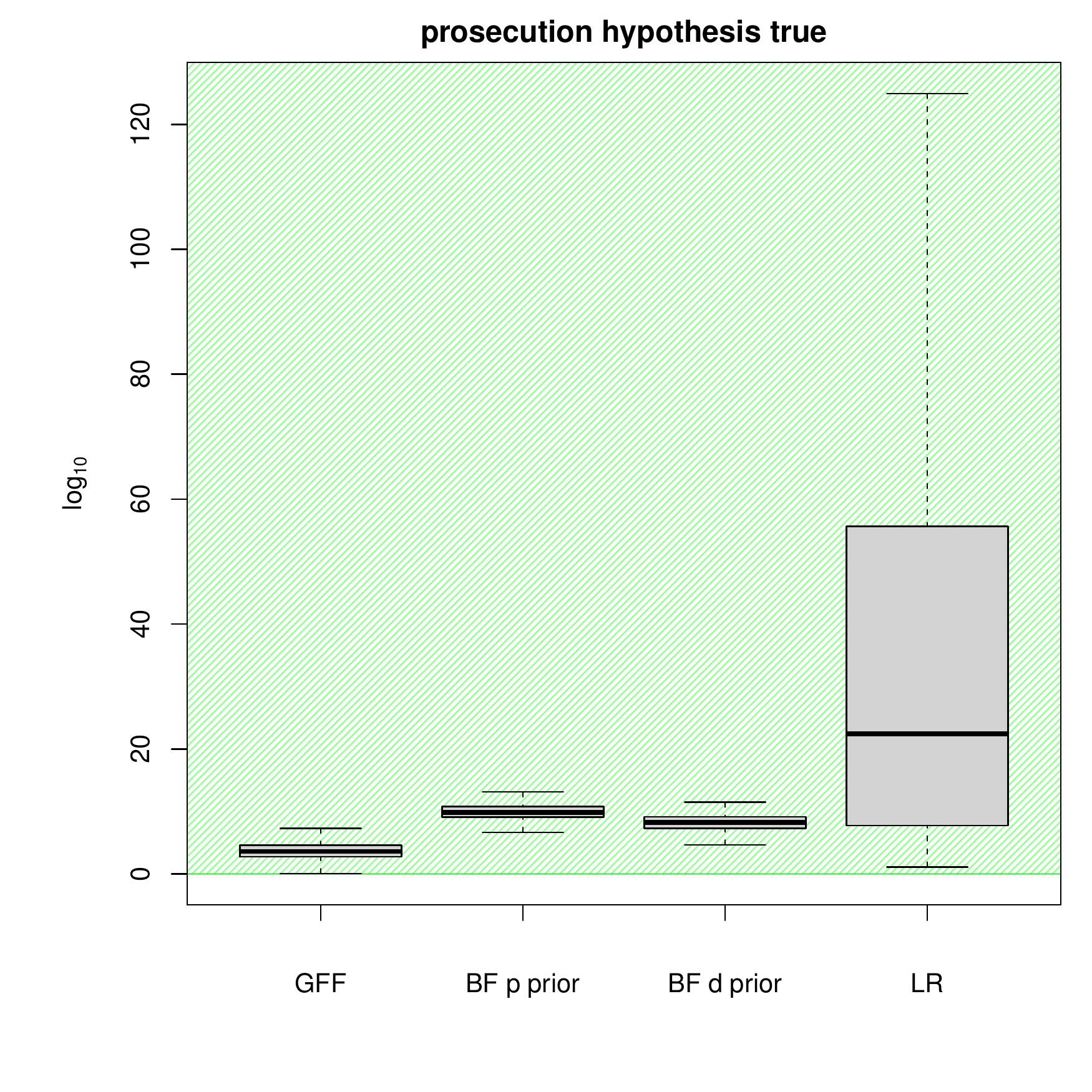}
\vspace{-.6in}\caption{\scriptsize Box plots of the sampling distributions of the GFF, BF, and LR over the 3,000 simulations under $H_{d}$ (left panel) and 320 simulations under $H_{p}$ (right panel).  For this synthetic `ideal sample size' simulation, $m_{u} = 2$, $m = 150$, $n = 659$, and $m_{i} = 3$.  BF p prior denotes the BF constructed from priors that favor $H_{p}$, whereas BF d prior denotes the BF constructed from priors that favor $H_{d}$.  The shaded green regions in each panel correspond to values of the GFF, BF, and LR that favor the true hypothesis. Outliers are omitted.}\label{boxplots_synthetic_ideal}
\vspace{-.15in}
\end{wrapfigure}

First, Figure \ref{boxplots_synthetic_ideal} presents a box plot of the performance of the GFF, BF, and LR over the 3,000 simulations under $H_{d}$ and 320 simulations under $H_{p}$.  The BF is evaluated both with a prior that favors the prosecution hypothesis, denoted `BF p prior', and with a prior that favors the defense hypothesis, denoted `BF d prior'.  The construction of these competing priors is as described in the previous section, where the prior specification is discussed.  Figure \ref{boxplots_synthetic_ideal} demonstrates that all four methods perform as reasonably desired in this ideal size synthetic data simulation (i.e., their sampling distributions favor the true hypothesis in under either scenario).  Note that the arguably inconsequential difference in the performance of the BF p prior versus BF d prior is a result of the unrealistically ideal sample sizes of this synthetic simulation design.  The next simulation design illustrates this point.

\begin{wrapfigure}{r}{0.4\textwidth}
\vspace{-.15in}
\centering
\includegraphics[scale=.35]{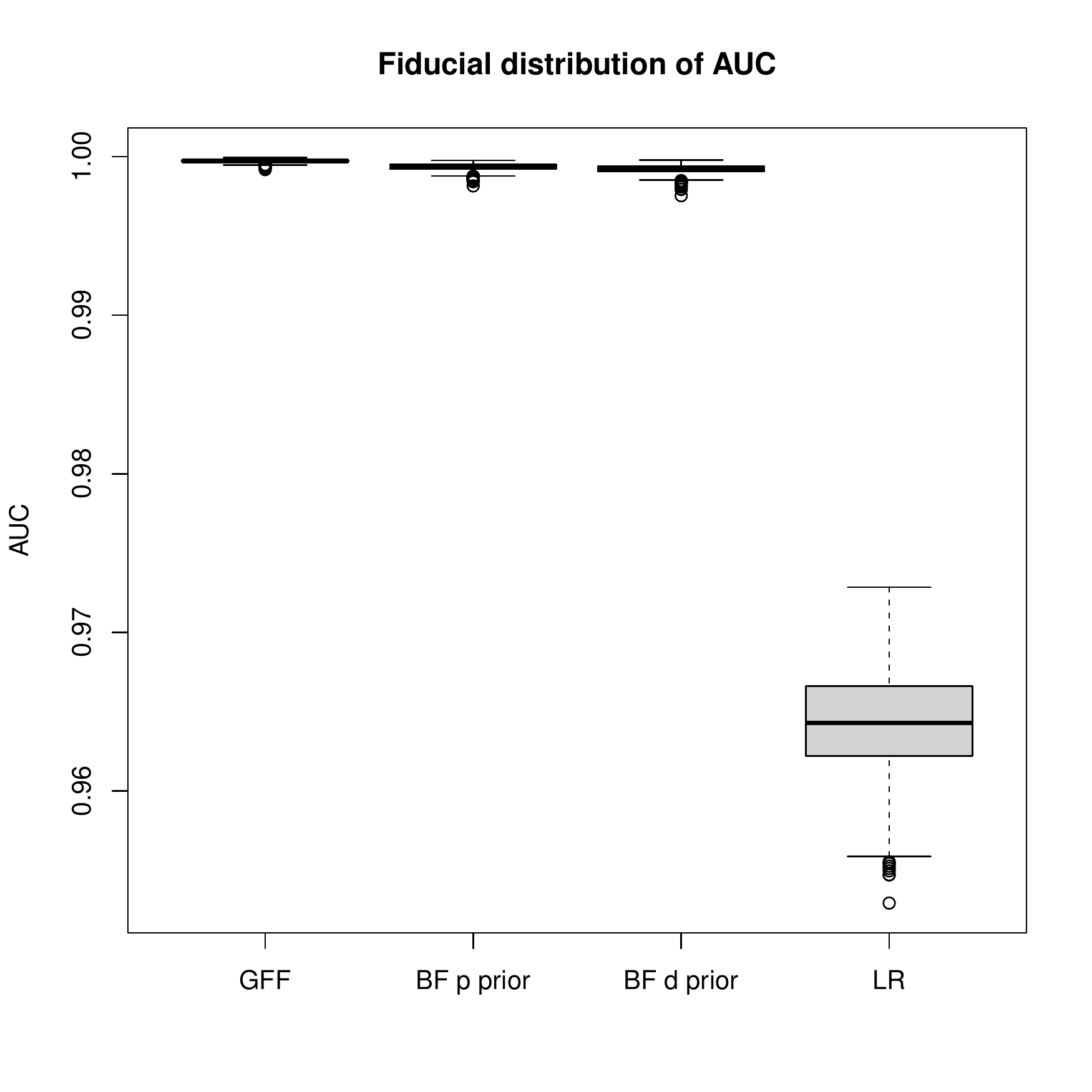}
\vspace{-.3in}\caption{\scriptsize Fiducial distributions of the AUC for the GFF, BF, and LR over the 3,000 simulations under $H_{d}$ and 320 simulations under $H_{p}$.  For this `ideal sample size' simulation, $m_{u} = 2$, $m = 150$, $n = 659$, and $m_{i} = 3$.}\label{auc_synthetic_ideal}
\vspace{-.15in}
\end{wrapfigure}

Second, the fiducial distributions of the area under the receiver operating characteristic curve (AUC) for the GFF, BF, and LR are displayed in Figure \ref{auc_synthetic_ideal}.  The AUC measures the adequacy of each of the four methods for accurately discriminating between $H_{p}$ and $H_{d}$, and the observed AUC values reflect an important feature observed in the distributions of the GFF, BF, and LR values in Figure \ref{boxplots_synthetic_ideal}.  There is almost no overlap in the observed GFF, BF p prior, and BF d prior values, respectively, for $H_{d}$ true versus $H_{p}$ true, which means there exist an effective threshold for discriminating between these two hypotheses for each of these methods.  Hence, the AUC values are clustered very close to the boundary at one in Figure \ref{auc_synthetic_ideal}.  Conversely, the LR values exhibit some overlap in tail values between $H_{d}$ true versus $H_{p}$ true, and so the LR AUC values reflect this loss of discriminating ability, though not a dramatic loss in this ideal sized simulation design.

Next, a meaningful notion for assessing the performance of ratio quantities such as the GFF, BF, and LR is to determine whether they are well calibrated to the values they exhibit.  For example, an LR value of 3 has the interpretation that it is 3 times as likely to observe the evidence when $H_{p}$ is true than when $H_{d}$ is true.  For this interpretation to be meaningful, for every instance that we observe an LR of 3 when $H_{d}$ is true we should observe 3 instances of an LR of 3 when $H_{p}$ is true.  As described in \cite{hannig2019}, a shorthand for this notion of calibration is the the expression `LR(LR) = LR'.  

\begin{figure}[H]
\vspace{-.15in}
\centering
\includegraphics[scale=.35]{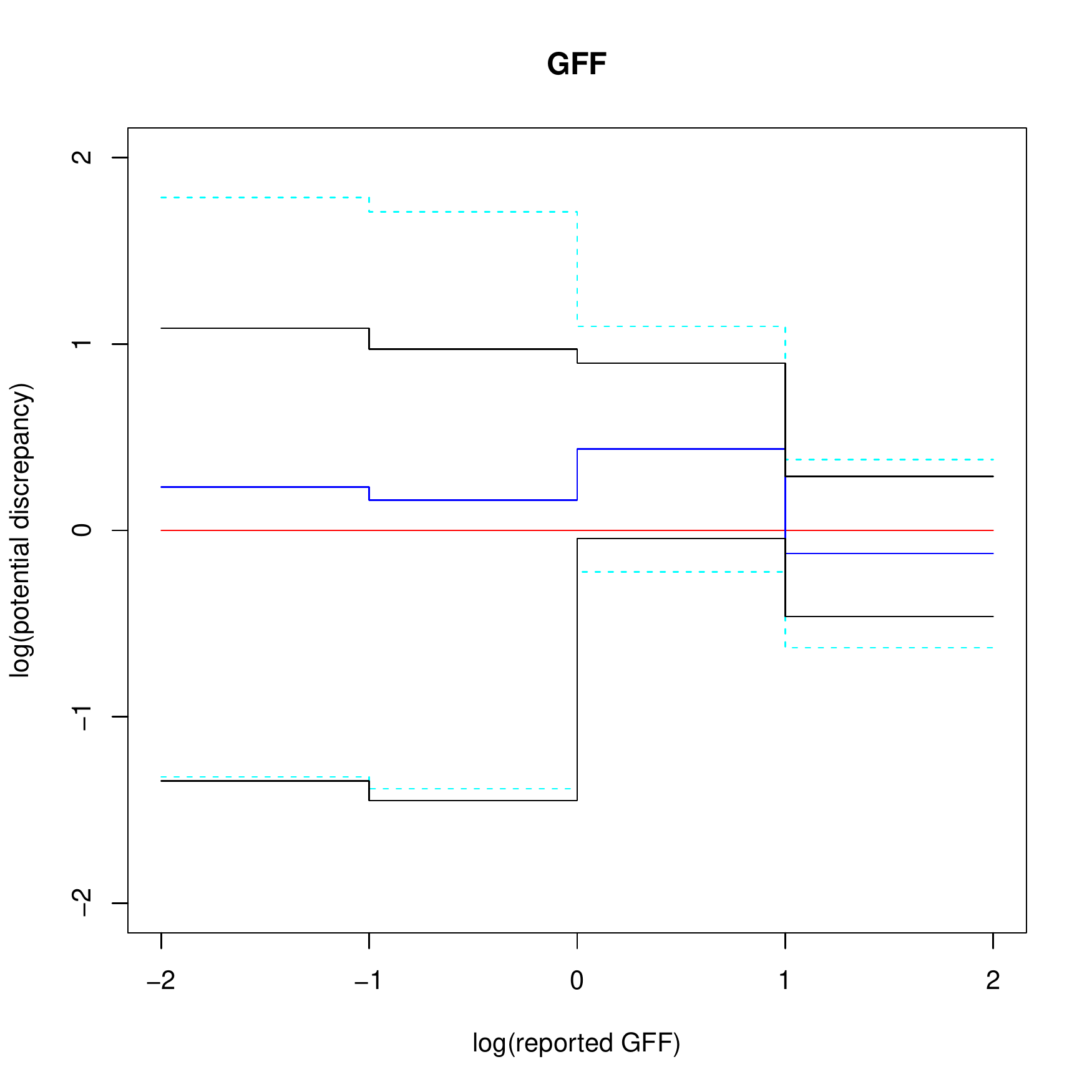}\includegraphics[scale=.35]{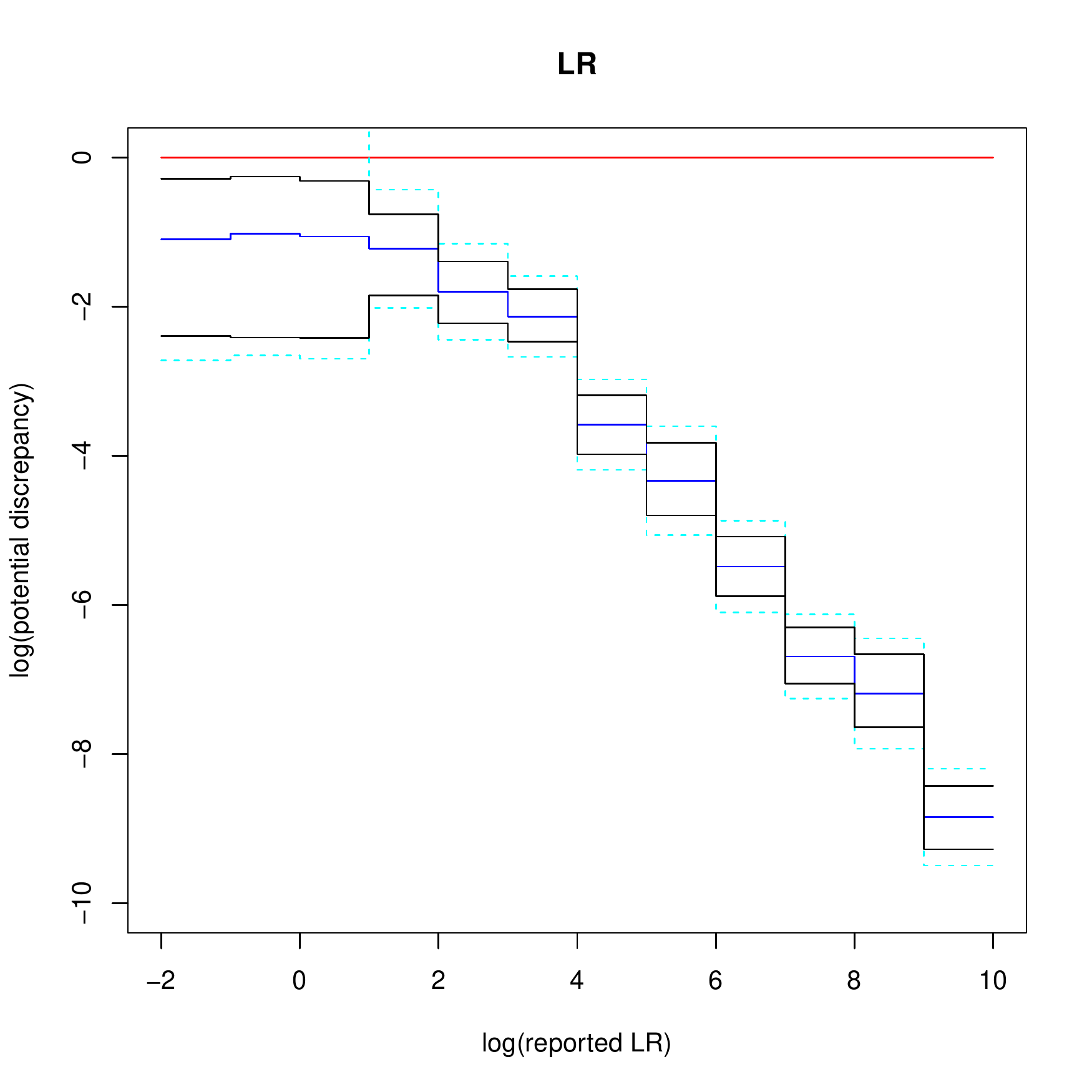}
\includegraphics[scale=.35]{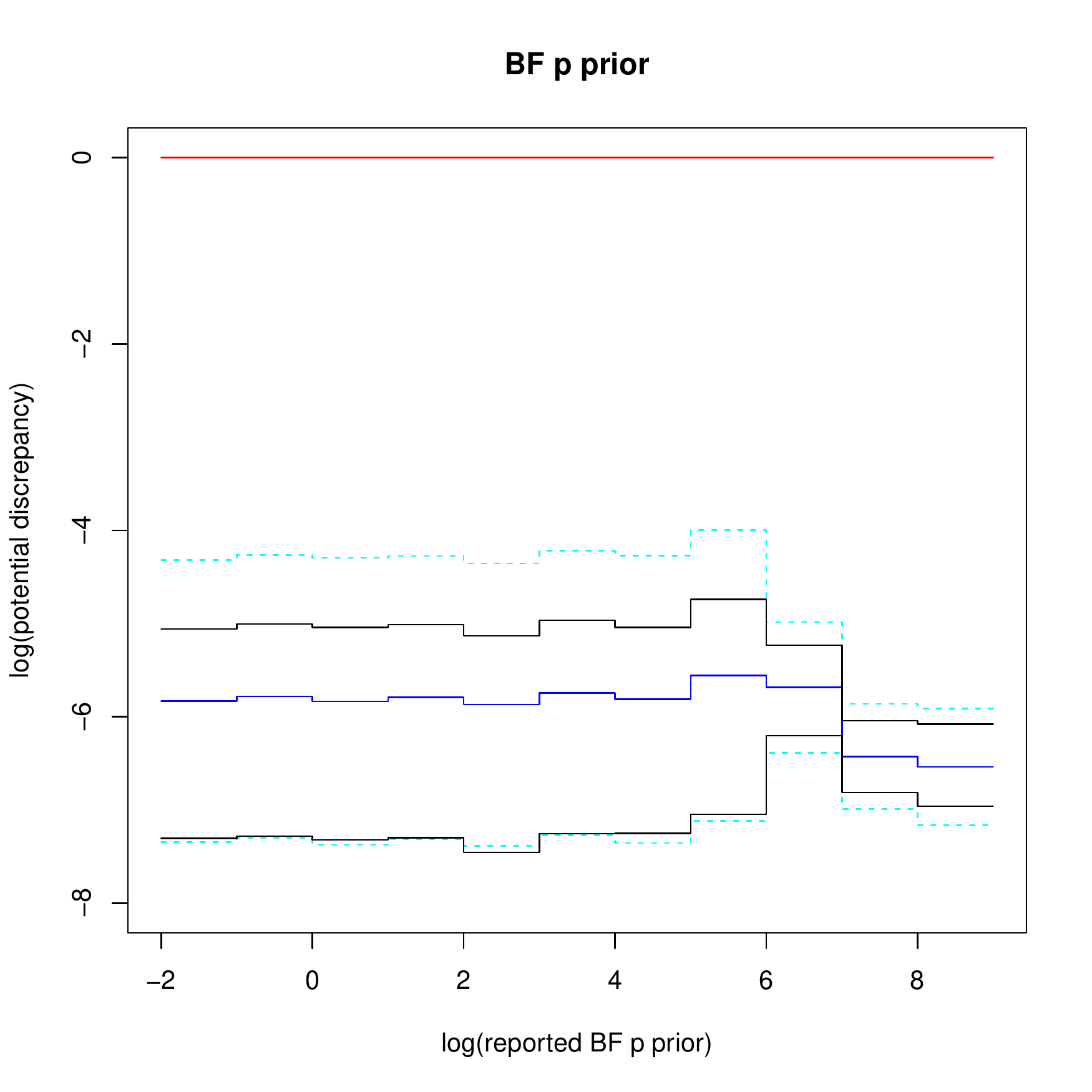}\includegraphics[scale=.35]{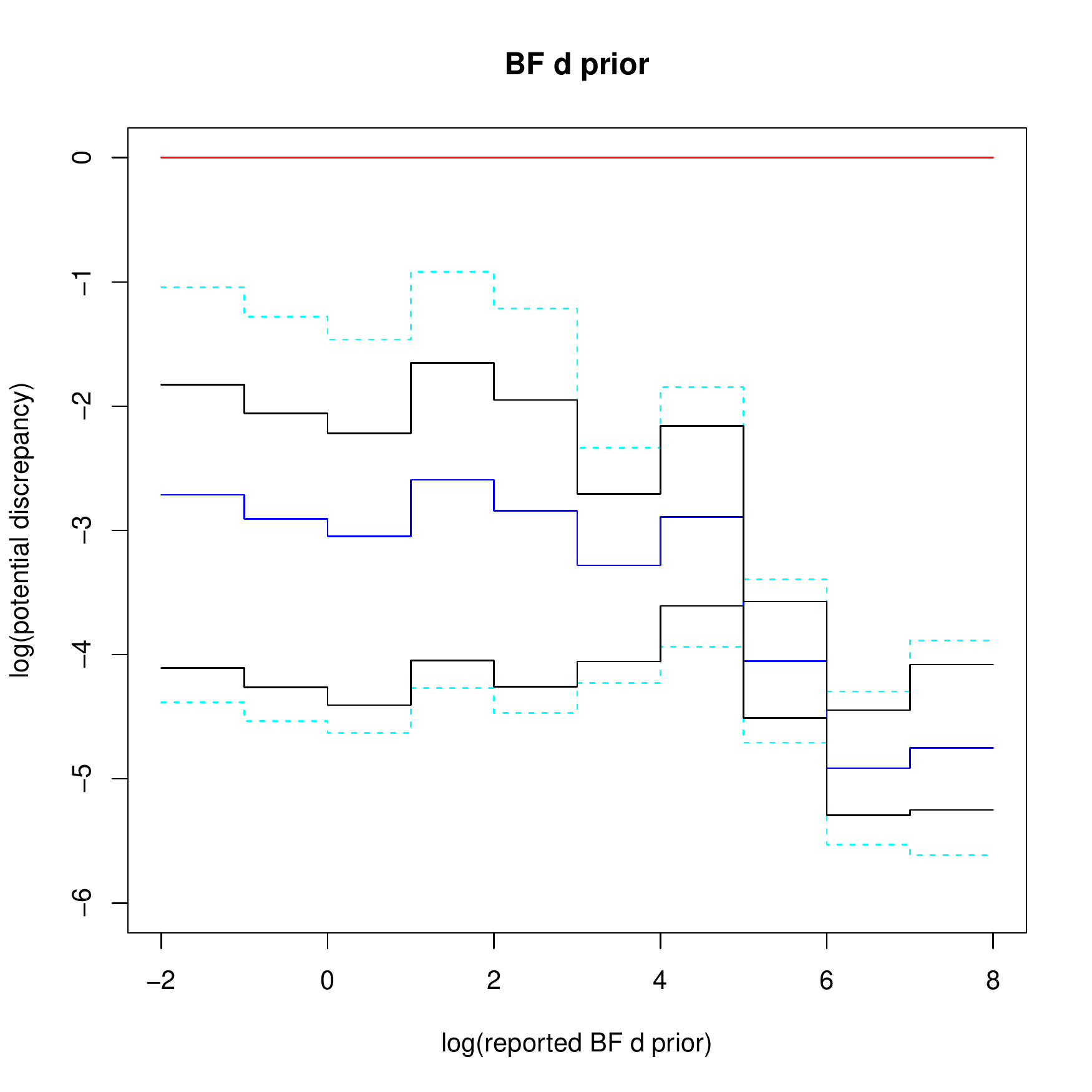}
\vspace{-.2in}\caption{\scriptsize Calibration for the GFF, BF, and LR over the 3,000 simulations under $H_{d}$ and 320 simulations under $H_{p}$.  The horizontal red line at zero corresponds to perfect calibration (i.e., LR(LR) = LR).  The blue line is the fiducial median log discrepancy.  The black and cyan lines are upper and lower .95 point-wise and simultaneous fiducial confidence intervals, respectively, for the log discrepancy.  For this `ideal sample size' simulation, $m_{u} = 2$, $m = 150$, $n = 659$, and $m_{i} = 3$.}\label{calibration_synthetic_ideal}
\vspace{-.15in}
\end{figure}

\begin{figure}[H]
\vspace{-.15in}
\centering
\includegraphics[scale=.35]{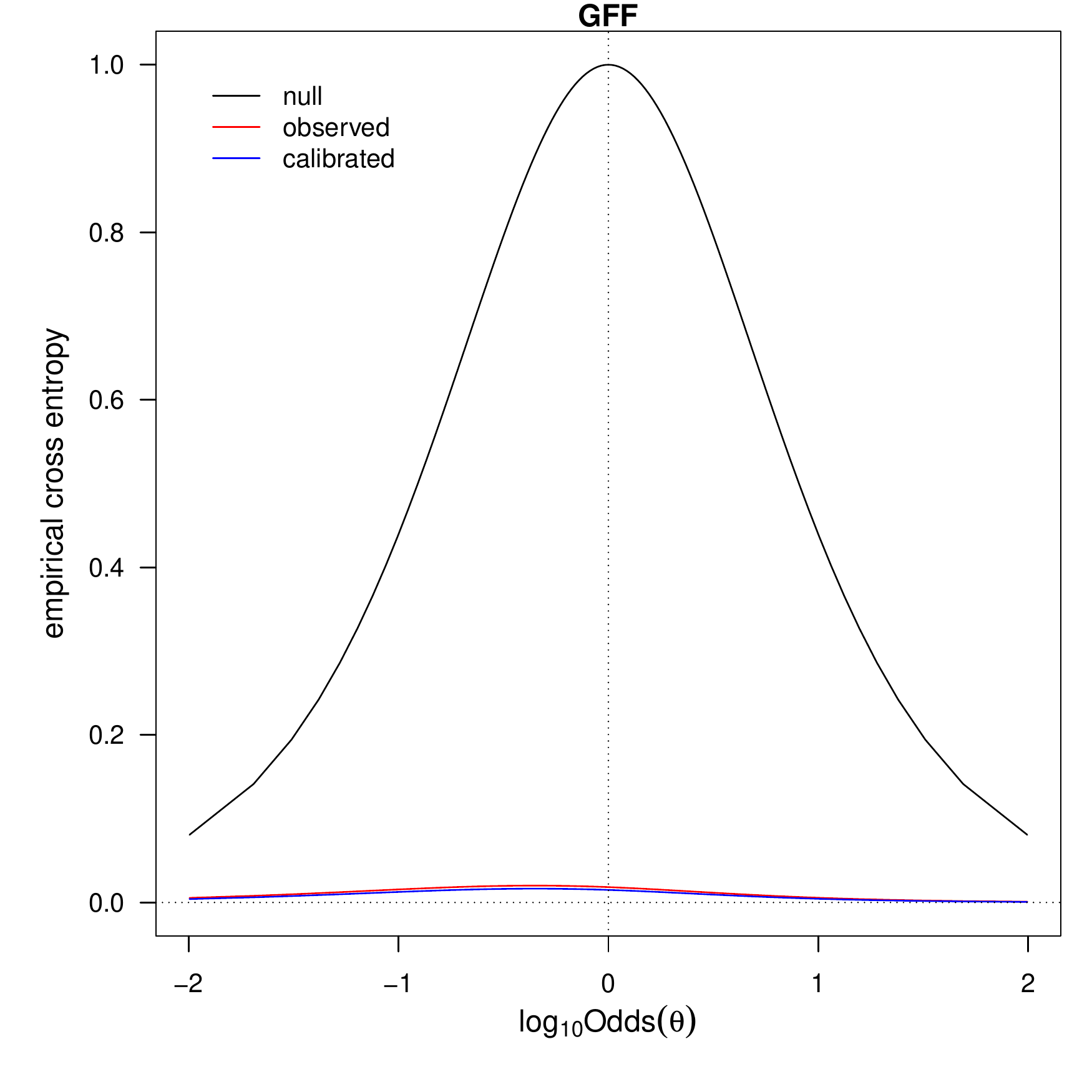}\includegraphics[scale=.35]{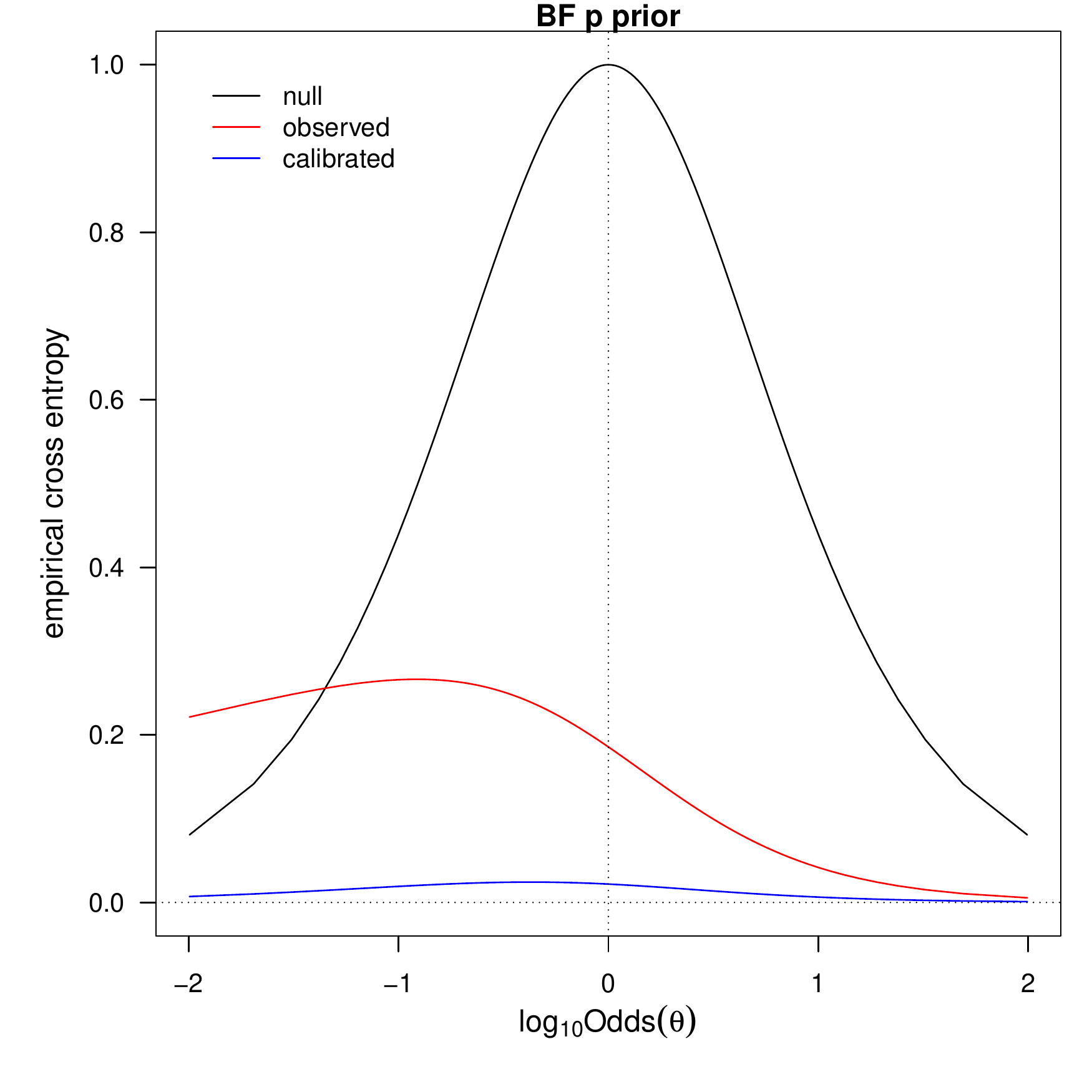}\includegraphics[scale=.35]{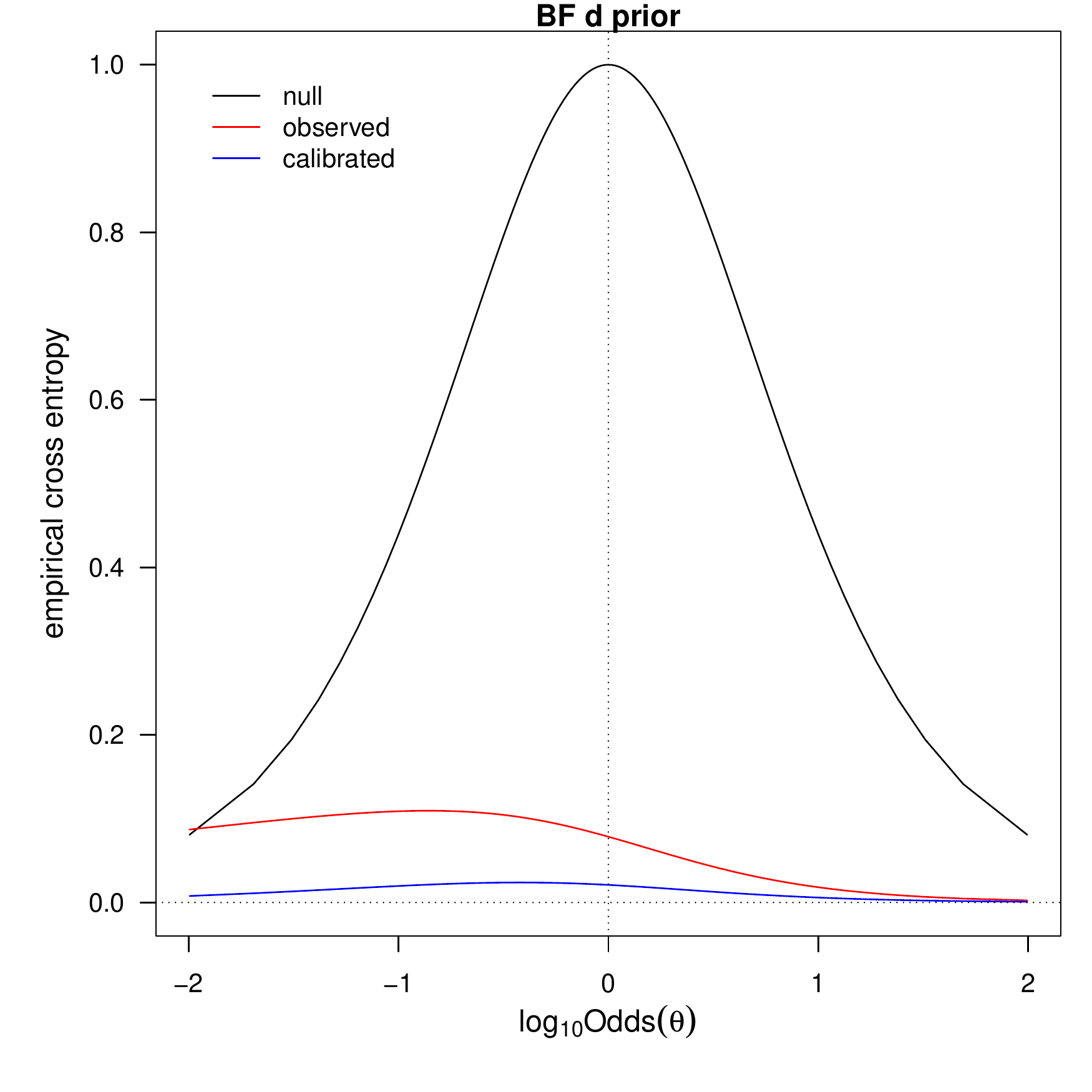}
\vspace{-.5in}\caption{\scriptsize Empirical cross entropy for the GFF, BF, and LR over the 3,000 simulations under $H_{d}$ and 320 simulations under $H_{p}$.  This calibration diagnostic tool is proposed in \cite{ramos2008}.  Good calibration is exhibited when the red line is nested between the blue and black lines, and as close as possible the blue.  The code from \cite{ramos2008} crashed for the LR, and for all subsequent simulation designs.  For this `ideal sample size' simulation, $m_{u} = 2$, $m = 150$, $n = 659$, and $m_{i} = 3$.}\label{ece_synthetic_ideal}
\vspace{-.15in}
\end{figure}

We follow the method in \cite{hannig2019} for estimating the calibration of the 3,000 simulations under $H_{d}$ and 320 simulations under $H_{p}$, for GFF, BF, and LR.  See Figure \ref{calibration_synthetic_ideal} for the estimated calibrations, and observe that the GFF is the best calibrated of the four methods.  Note that these ratio quantities can yield very poorly calibrated values while still being effective at discriminating between hypotheses, as seen for the LR, BF p prior, and the BF d prior.  The consequence of poor calibration is a misrepresentation, often an exaggeration of the strength of evidence supporting the respective hypotheses.  In the context of forensic identification of source problems, such misrepresentation can lead to the false conclusion that the evidence in favor of a particular hypothesis is overwhelming, or beyond any doubt.  Thus, the implication of a lack of calibration cannot be overstated.

We conclude this section by presenting an alternative calibration analysis described in \cite{ramos2008}.  See Figure \ref{ece_synthetic_ideal}.  It is again observed that the GFF values are the best calibrated.  Unfortunately, the code \citep{lucy2013} for this calibration analysis only worked for the GFF, BF p prior, and BF d prior values in this ideal size synthetic data simulation, and so similar figures are not available for the two simulation designs that follow.

\subsection{Simulation 2: fully synthetic data with NFI data sample sizes}\label{simulation2}

\begin{wrapfigure}{r}{0.75\textwidth}
\vspace{-.15in}
\centering
\includegraphics[scale=.35]{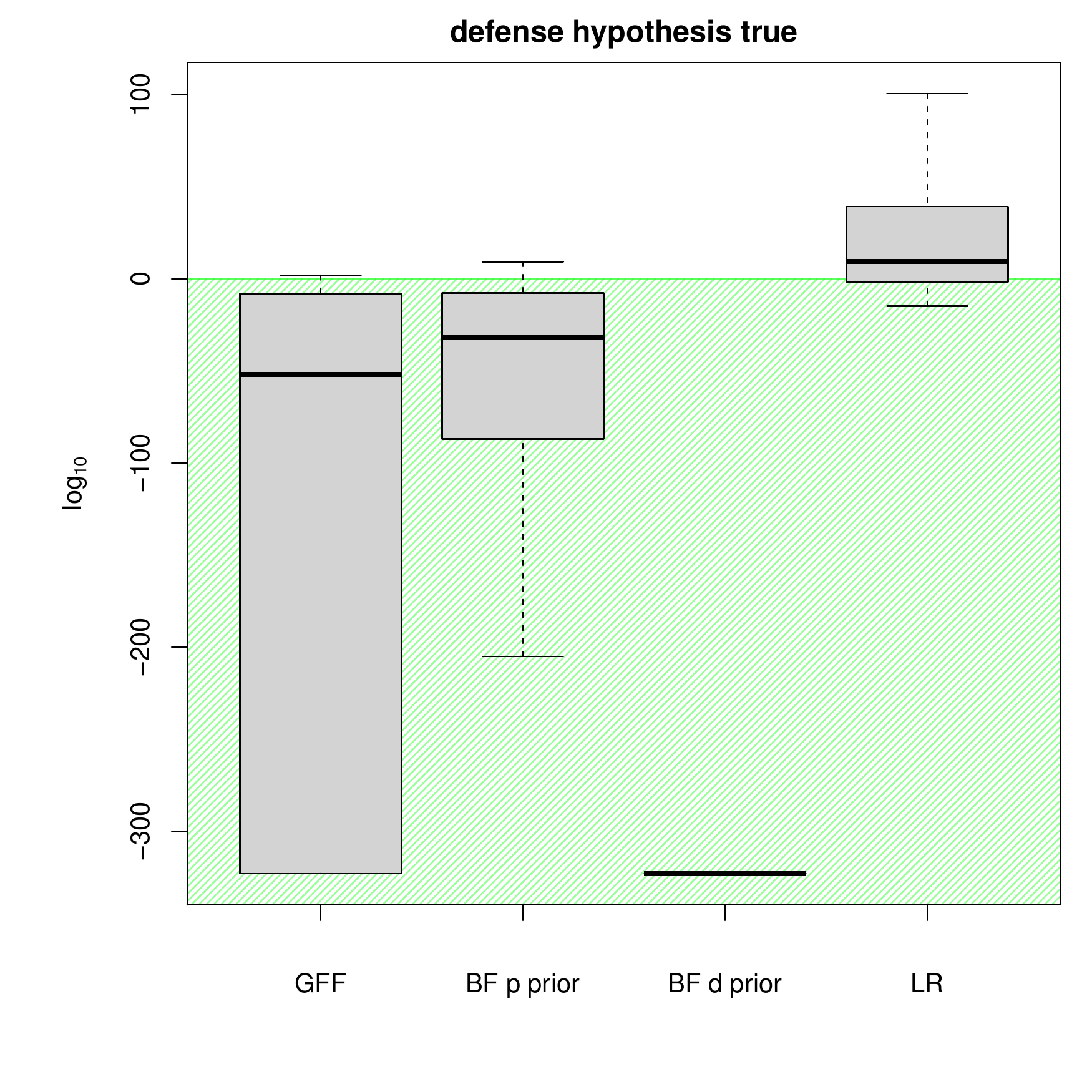}\includegraphics[scale=.35]{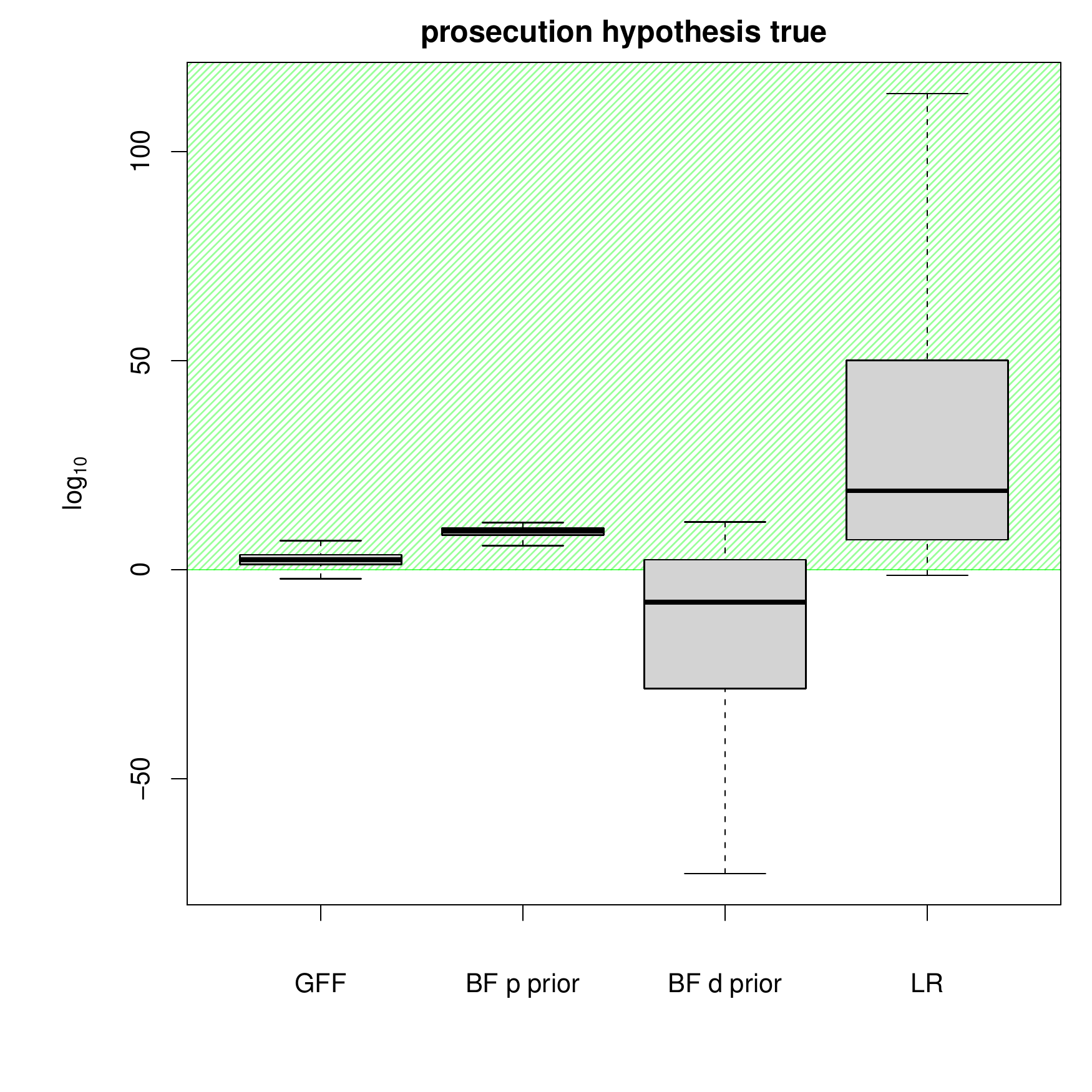}
\vspace{-.6in}\caption{\scriptsize Box plots of the sampling distributions of the GFF, BF, and LR over the 3,000 simulations under $H_{d}$ (left panel) and 320 simulations under $H_{p}$ (right panel).  For this synthetic `NFI casework data sample sizes' simulation, $m_{u} = 2$, $m = 3$, $n = 659$, and $m_{i} = 3$.  BF p prior denotes the BF constructed from priors that favor $H_{p}$, whereas BF d prior denotes the BF constructed from priors that favor $H_{d}$.  The shaded green regions in each panel correspond to values of the GFF, BF, and LR that favor the true hypothesis.  Outliers are omitted.}\label{boxplots_synthetic_actual}
\vspace{-.15in}
\end{wrapfigure}

The sampling distributions are displayed in Figure \ref{boxplots_synthetic_actual}.  A first observation is that the LR tends to favor $H_{p}$ in both scenarios, and as noted for the previous simulation design this results from an unstable MLE of the specific source parameters with $m = 3$.  Referring back to the LR construction in equation (\ref{LR_computation}), the instability stems from the evaluation of $f_{s}\big(\{y_{s,k}\} \mid \widehat{\theta}_{s}\big)$ in the denominator.

\begin{wrapfigure}{r}{0.4\textwidth}
\vspace{-.15in}
\centering
\includegraphics[scale=.35]{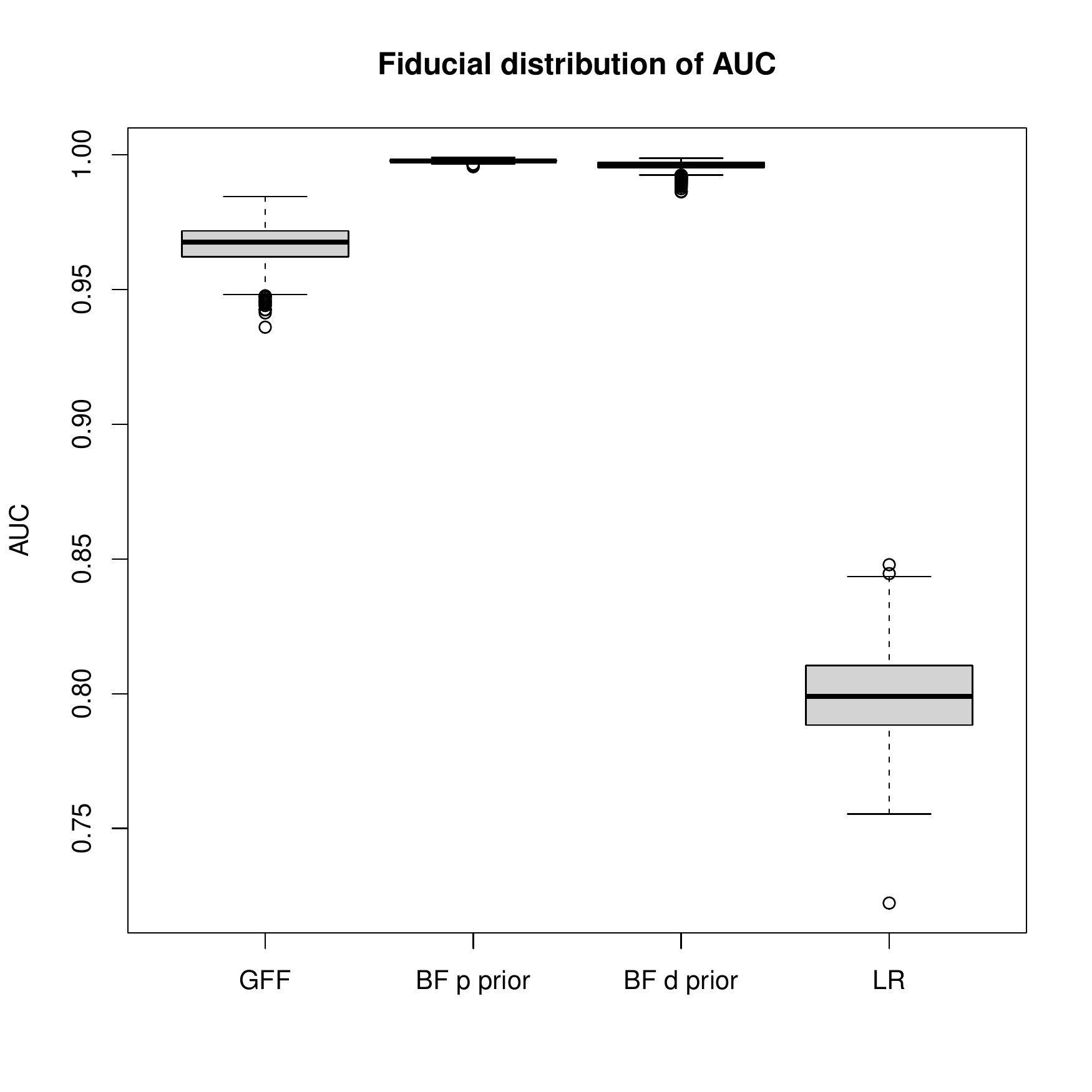}
\vspace{-.3in}\caption{\scriptsize Fiducial distributions of the AUC for the GFF, BF, and LR over the 3,000 simulations under $H_{d}$ and 320 simulations under $H_{p}$.  For this synthetic `NFI casework data sample sizes' simulation, $m_{u} = 2$, $m = 3$, $n = 659$, and $m_{i} = 3$.}\label{auc_synthetic_actual}
\vspace{-.15in}
\end{wrapfigure}

\begin{figure}[H]
\vspace{-.15in}
\centering
\includegraphics[scale=.35]{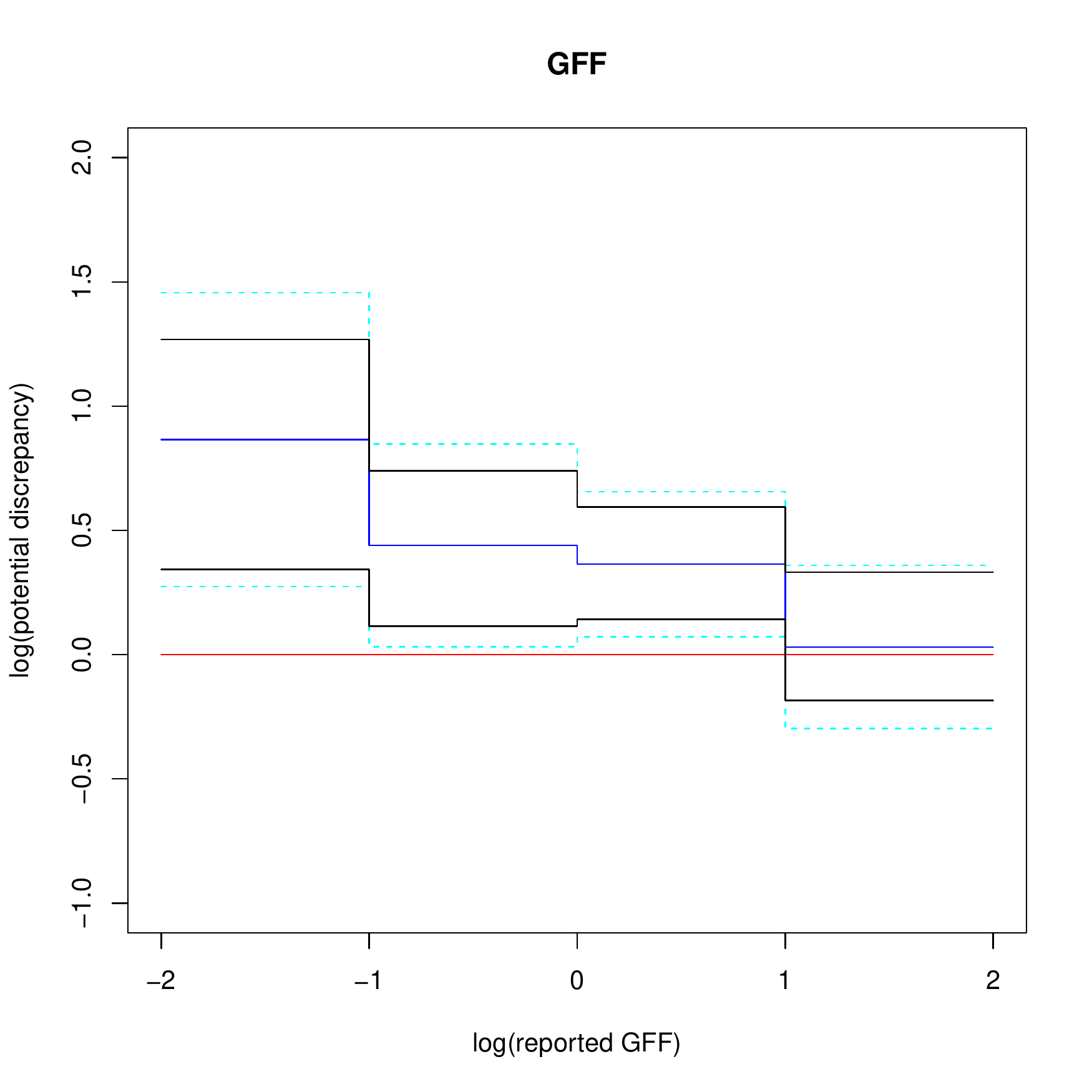}\includegraphics[scale=.35]{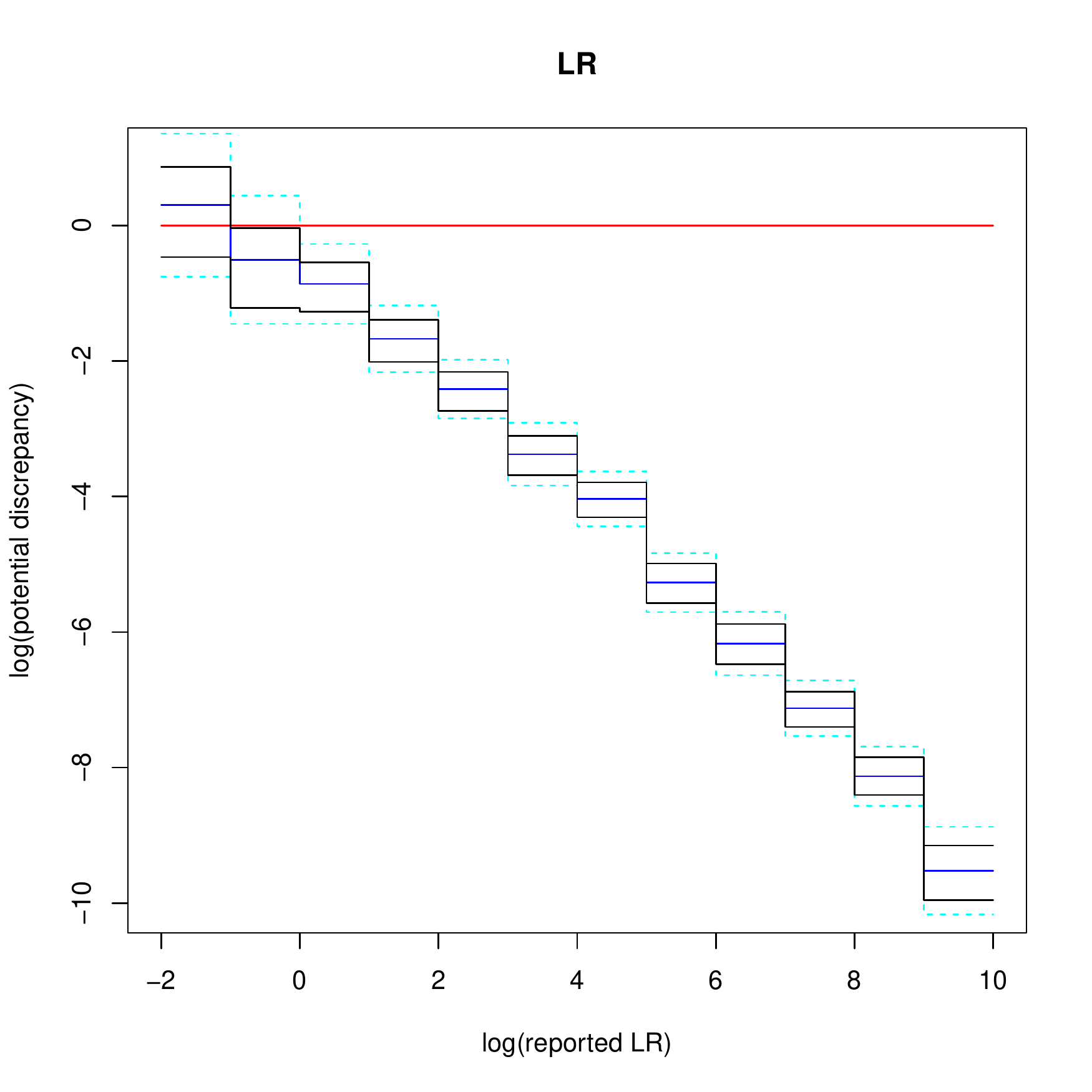}\includegraphics[scale=.35]{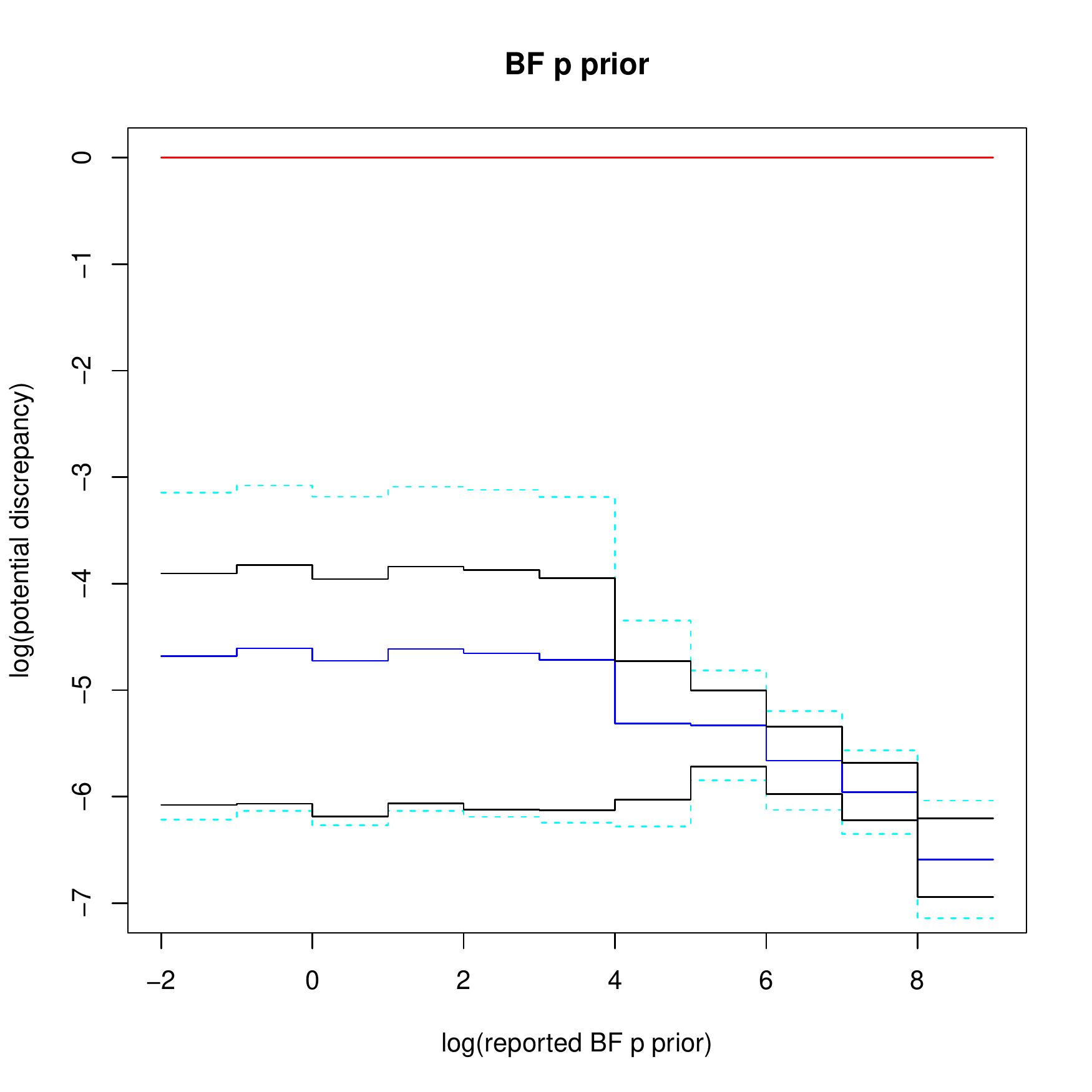}
\vspace{-.45in}\caption{\scriptsize Calibration for the GFF, BF, and LR over the 3,000 simulations under $H_{d}$ and 320 simulations under $H_{p}$.  The horizontal red line at zero corresponds to perfect calibration (i.e., LR(LR) = LR).  The blue line is the fiducial median log discrepancy.  The black and cyan lines are upper and lower .95 pointwise and simultaneous fiducial confidence intervals, respectively, for the log discrepancy.  For this `NFI casework data sample sizes' simulation, $m_{u} = 2$, $m = 3$, $n = 659$, and $m_{i} = 3$.}\label{calibration_synthetic_actual}
\vspace{-.15in}
\end{figure}

The next feature to observe in Figure \ref{boxplots_synthetic_actual} is that the strength of evidence for $H_{d}$ is characterized by the BF p prior an order of magnitude smaller than by the BF d prior, in the $H_{d}$ true scenario.  These prosecution and defense priors were constructed to reflect extreme beliefs, to demonstrate that any values between the BF p prior and BF d prior values can reasonably result from the prior specification.  The $H_{p}$ true scenario is even more problematic because the BF p prior and the BF d prior tend to favor opposite hypotheses.  This consequence of subjectivist Bayesian prior choice for forensic identification of source problems, as illustrated in Figure \ref{boxplots_synthetic_actual}, is exceedingly problematic because it demonstrates that the strength of evidence for or against a hypothesis is heavily influenced by the competing prior beliefs (prosecution versus defense) for or against the hypothesis, even to the point where the BF entirely favors the wrong hypothesis.  Conversely, it is observed in Figure \ref{boxplots_synthetic_actual} that the GFF values tend to favor the true hypothesis in each scenario.  Moreover, the GFF values do not suffer from the instability exhibited by the LR values.  These important features illustrated in Figure \ref{boxplots_synthetic_actual} are further supported by the discrimination and calibration analyses presented in Figures \ref{auc_synthetic_actual} and \ref{calibration_synthetic_actual}, respectively.

As alluded to in the discussion for the previous simulation design, even if the values of the GFF, BF, or LR do not tend to be associated with the true hypothesis, it is still possible that these methods are effective at correctly discriminating between $H_{d}$ and $H_{p}$.  The most notable of these four methods is BF d prior values, as displayed in Figure \ref{boxplots_synthetic_actual}.  There is a clear distinction between the distribution of BF d prior values under $H_{d}$ versus $H_{p}$, even though both distributions tend to exhibit values associated with $H_{d}$ true.  The distinction between the BF d prior values for the two hypotheses is characterized by the fiducial AUC distributions shown in Figure \ref{auc_synthetic_actual} (along with that for the other three methods, as well).  As in the previous simulation design, with AUC values very close to one, the GFF, BF p prior, and BF d prior are very effective at discriminating between $H_{d}$ and $H_{p}$.  The LR suffers in its ability to discriminate, due to the issues with numerical instability for sample sizes so small, as described at the beginning of this section and illustrated in Figure \ref{boxplots_synthetic_actual}.

While in this `actual sample size' simulation design, the GFF and BF p prior methods tend to exhibit values associated with the correct hypothesis (i.e., Figure \ref{boxplots_synthetic_actual}) and are effective at discriminating between $H_{d}$ and $H_{p}$ (i.e., Figure \ref{auc_synthetic_actual}), there is still a danger that they are not calibrated to appropriately reflect the strength of evidence that their values suggest.  Figure \ref{calibration_synthetic_actual} presents the calibration analysis for the GFF, BF p prior, and LR values.  Note that the calibration for the BF d prior values is missing; the values are very poorly calibrated and so the calibration softwares crashed.  Furthermore, Figure \ref{calibration_synthetic_actual} suggests that the LR and BF p prior values are also poorly calibrated.  The GFF values are much better, and in fact, reasonably well calibrated in light of the very small sample sizes that characterize this simulation design and the real NFI casework data.

\subsection{Simulation 3: real NFI casework data}\label{simulation3}

\begin{wrapfigure}{r}{0.75\textwidth}
\vspace{-.15in}
\centering
\includegraphics[scale=.35]{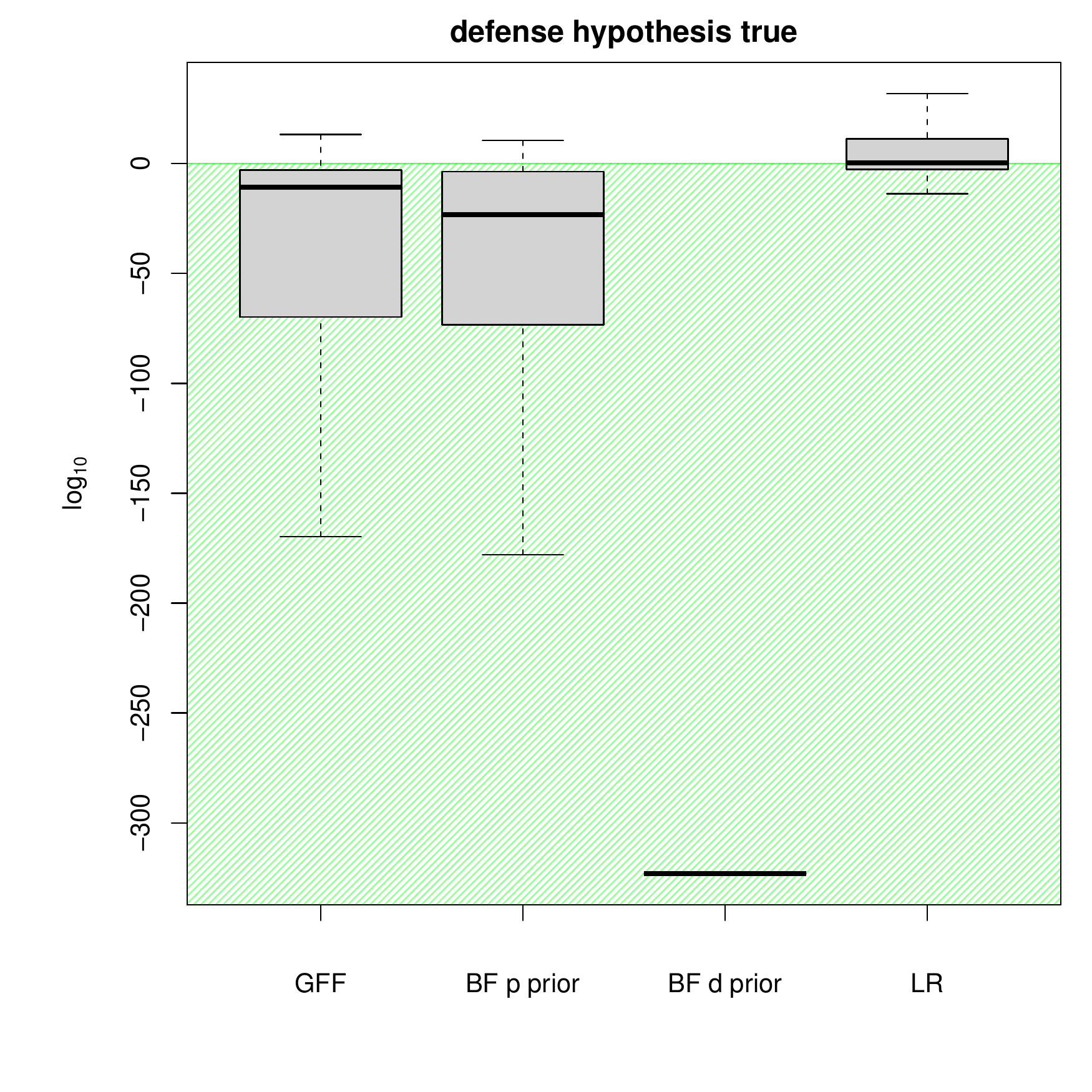}\includegraphics[scale=.35]{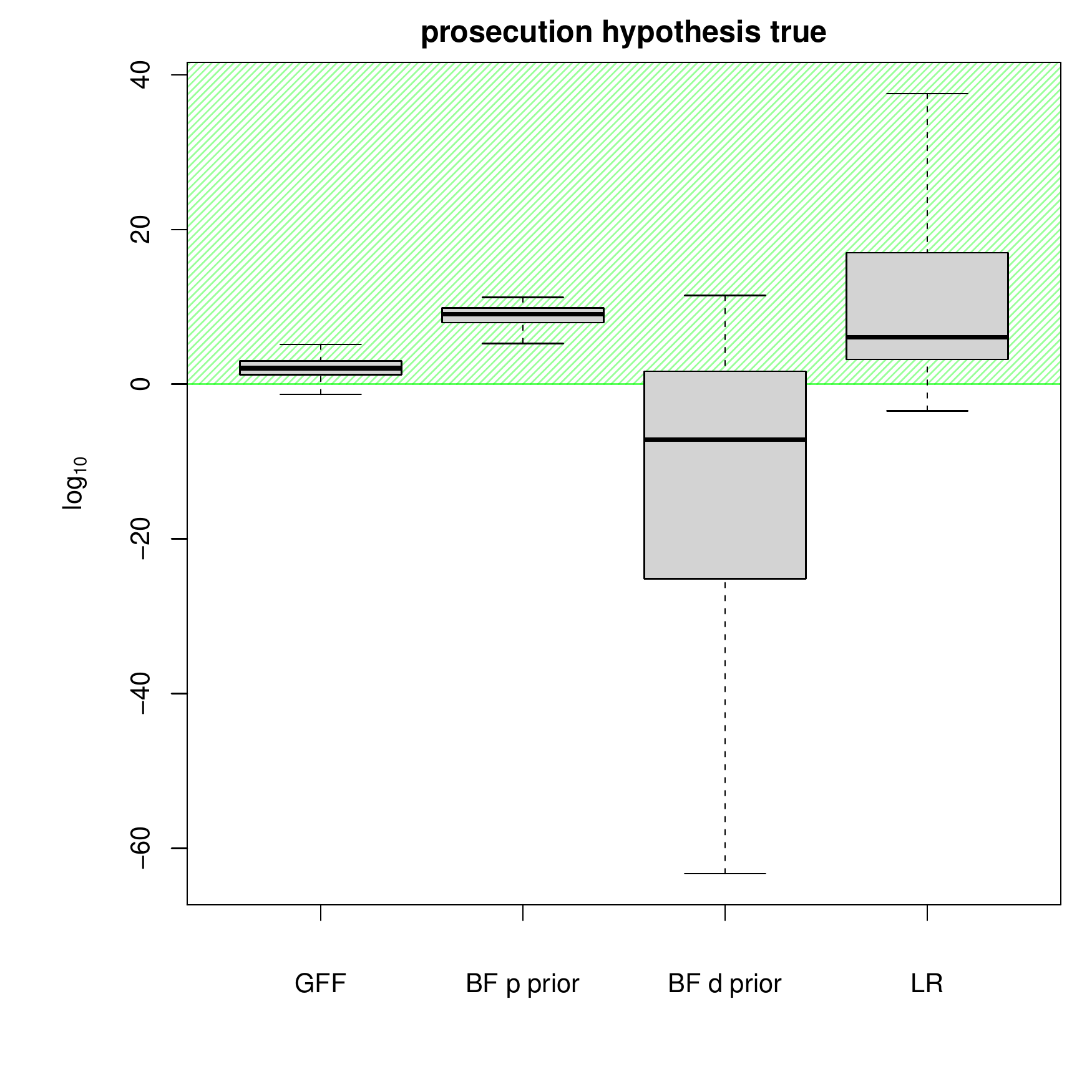}
\vspace{-.6in}\caption{\scriptsize Box plots of the sampling distributions of the GFF, BF, and LR over the 3,000 simulations under $H_{d}$ (left panel) and 320 simulations under $H_{p}$ (right panel).  For this `real NFI casework data' simulation, $m_{u} = 2$, $m = 3$, $n = 659$, and $m_{i} = 3$.  BF p prior denotes the BF constructed from priors that favor $H_{p}$, whereas BF d prior denotes the BF constructed from priors that favor $H_{d}$.  The shaded green regions in each panel correspond to values of the GFF, BF, and LR that favor the true hypothesis.  Outliers are omitted.}\label{boxplots_NFI_data}
\vspace{-.15in}
\end{wrapfigure}

\begin{wrapfigure}{r}{0.4\textwidth}
\vspace{-.15in}
\centering
\includegraphics[scale=.35]{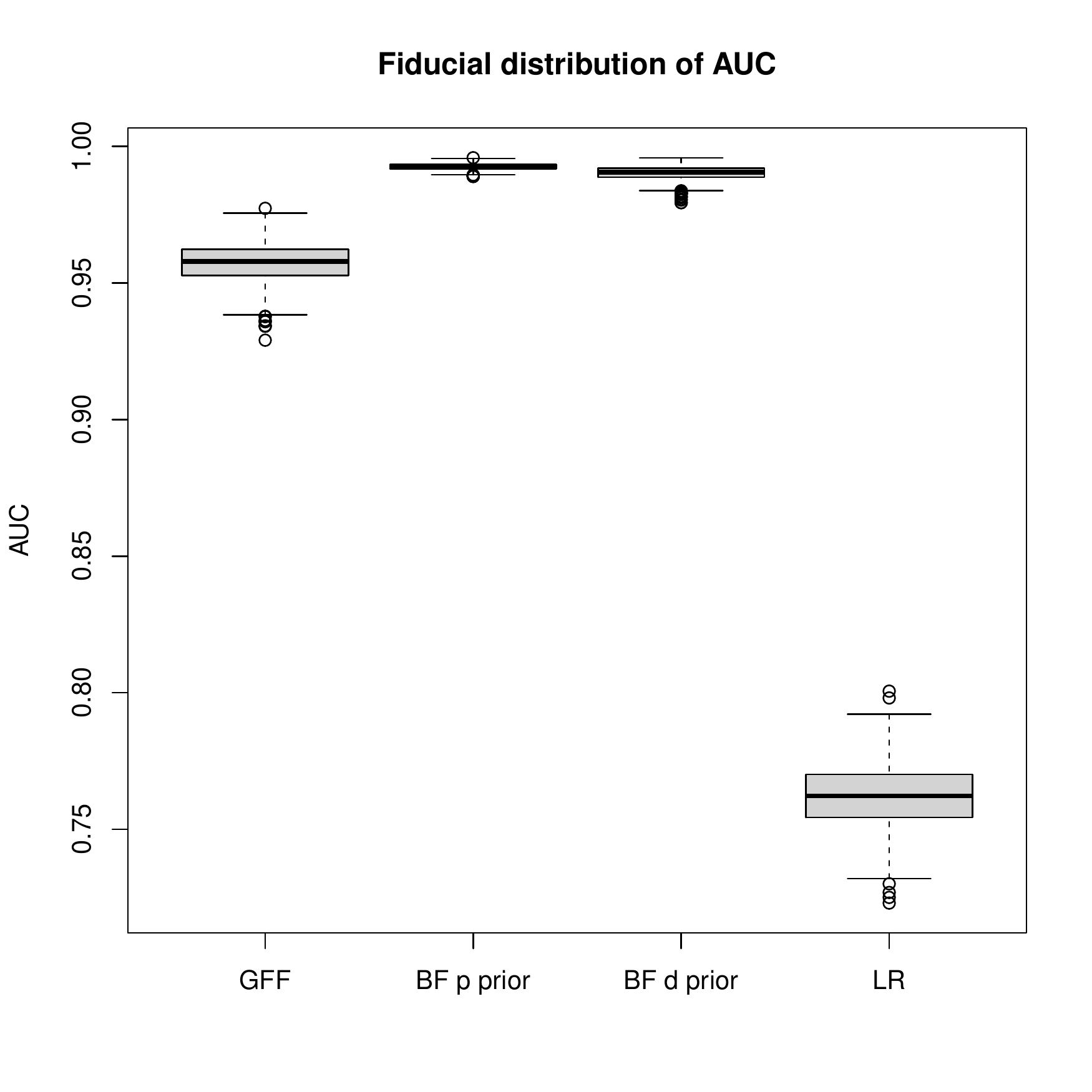}
\vspace{-.3in}\caption{\scriptsize Fiducial distributions of the AUC for the GFF, BF, and LR over the 3,000 simulations under $H_{d}$ and 320 simulations under $H_{p}$.  For this `real NFI casework data' simulation, $m_{u} = 2$, $m = 3$, $n = 659$, and $m_{i} = 3$.}\label{auc_NFI_data}
\vspace{-.15in}
\end{wrapfigure}

\begin{figure}[H]
\vspace{-.15in}
\centering
\includegraphics[scale=.35]{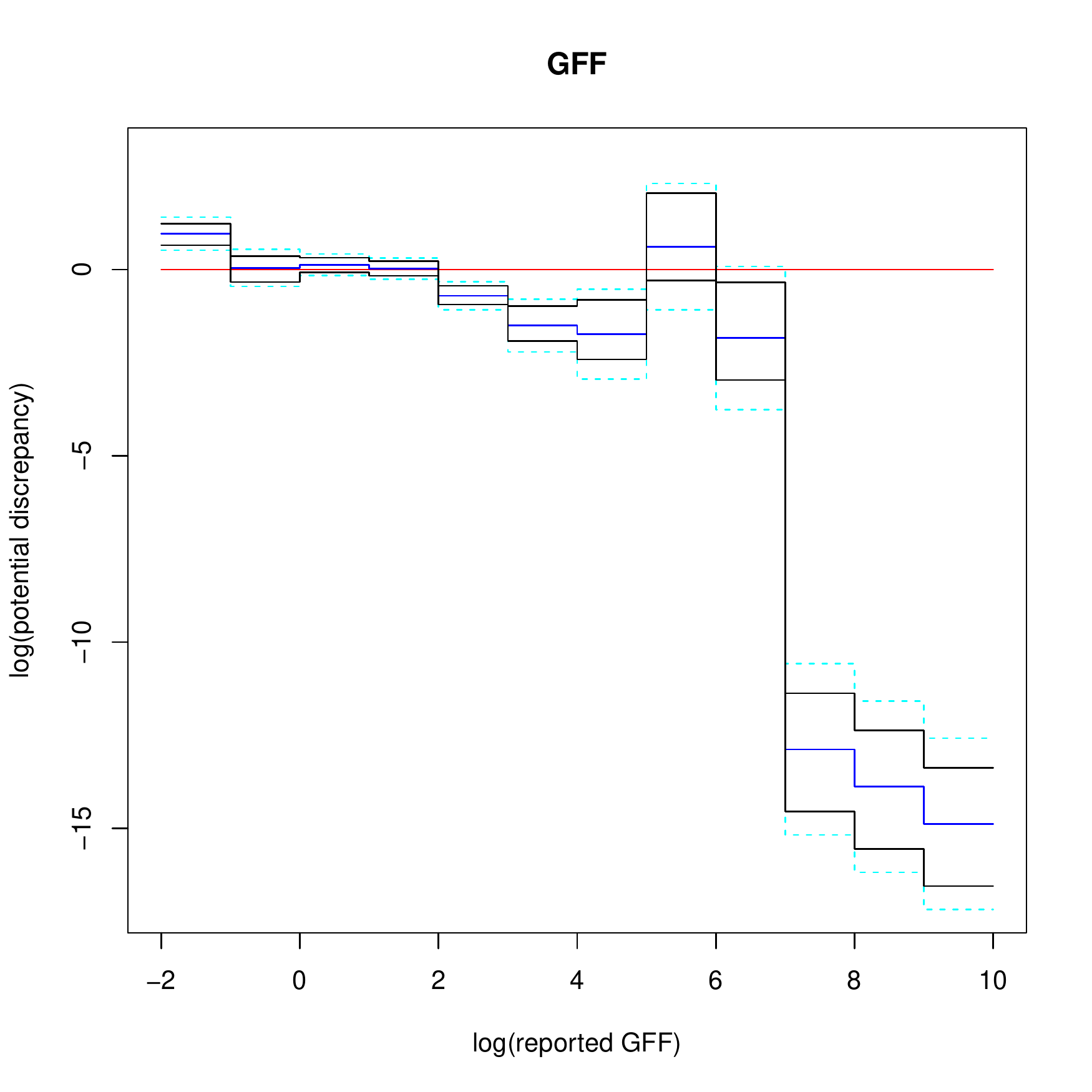}\includegraphics[scale=.35]{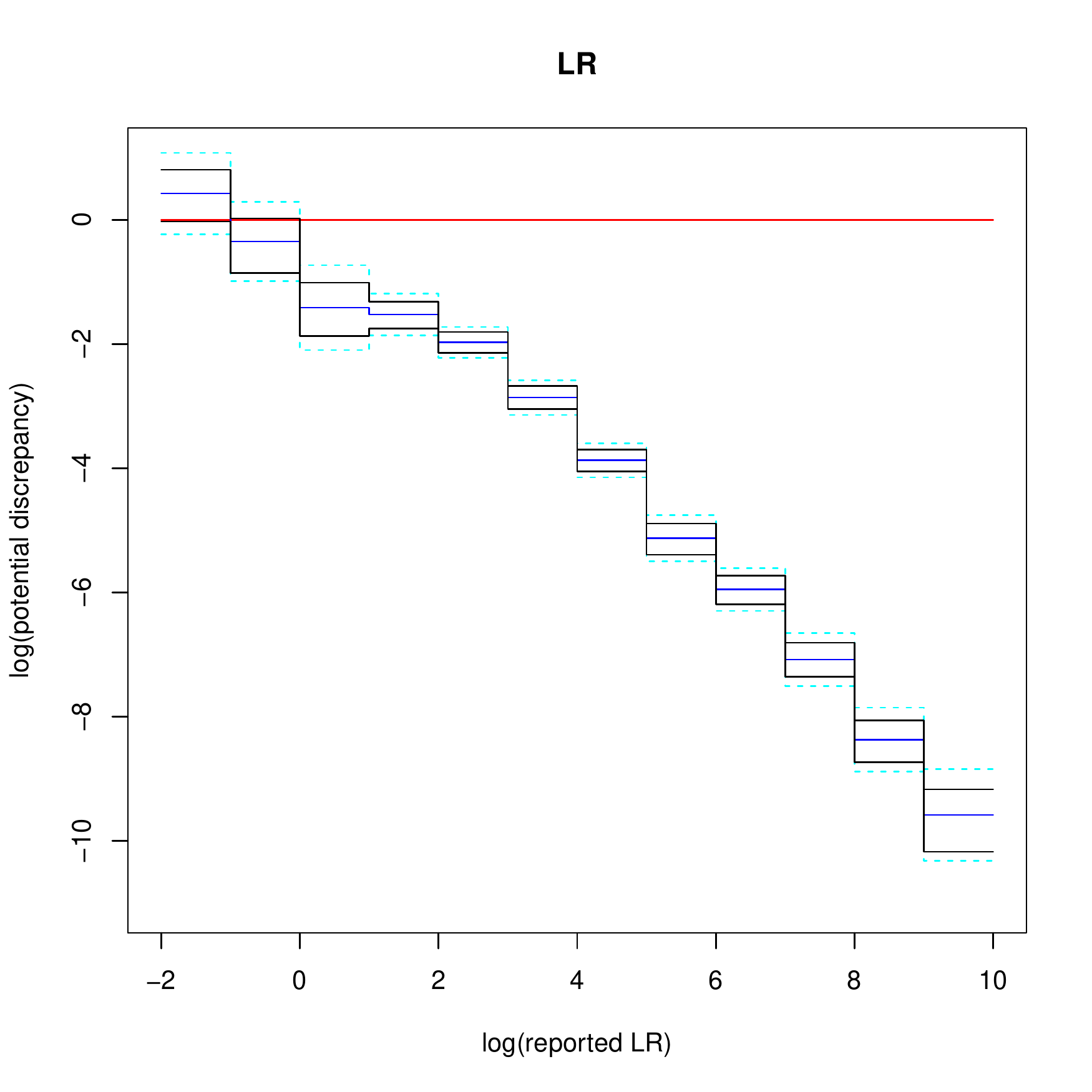}
\includegraphics[scale=.35]{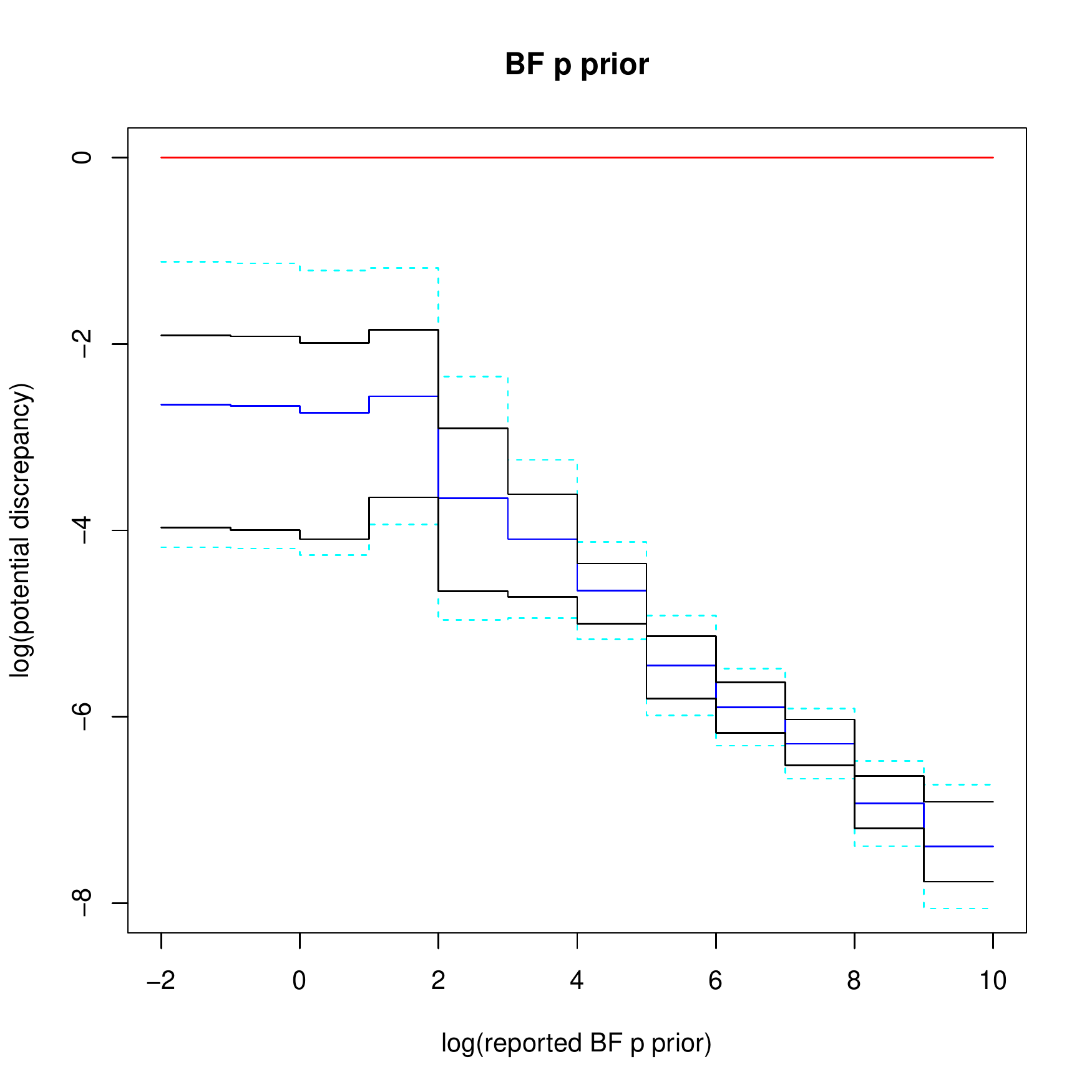}\includegraphics[scale=.35]{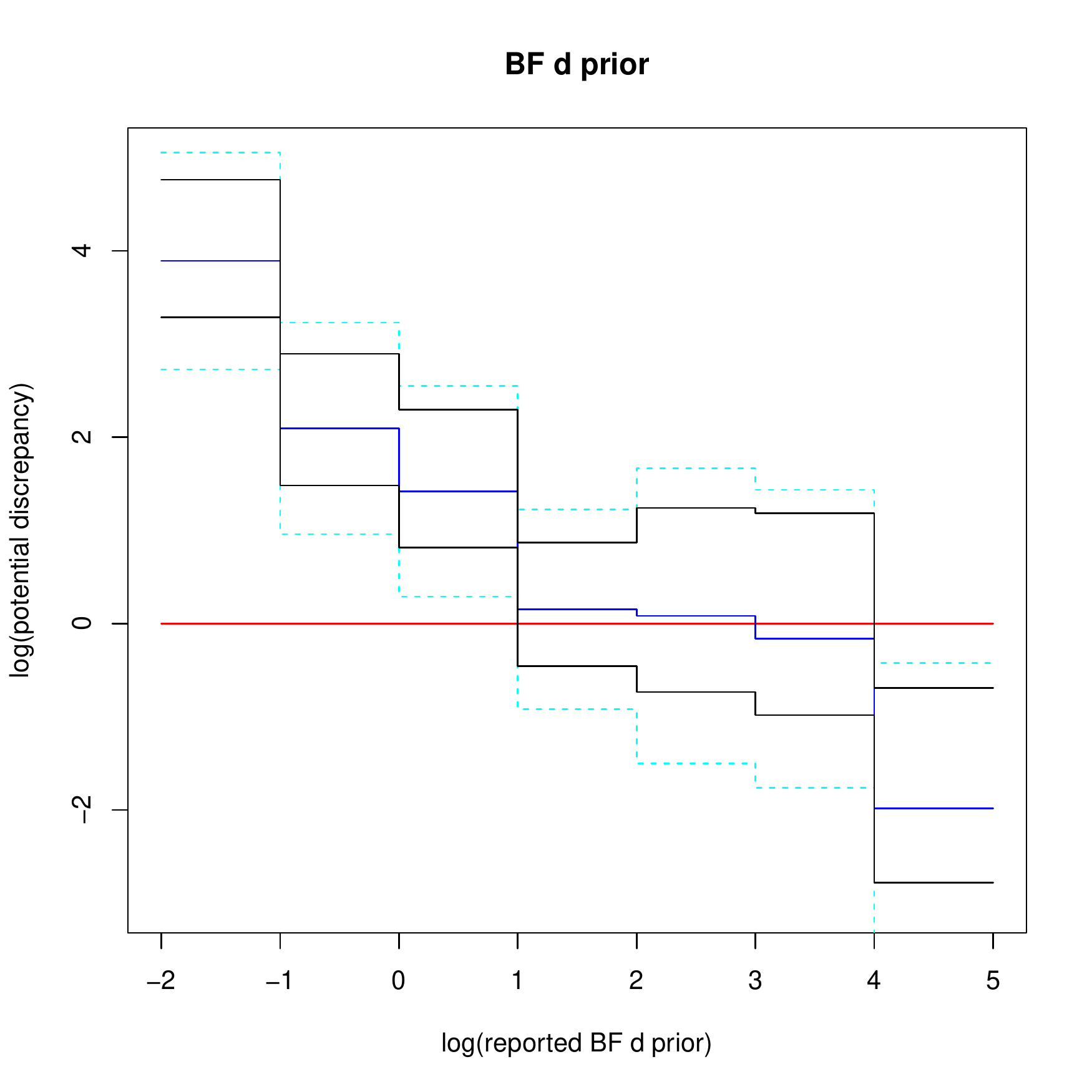}
\vspace{-.2in}\caption{\scriptsize Calibration for the GFF, BF, and LR over the 3,000 simulations under $H_{d}$ and 320 simulations under $H_{p}$.  The horizontal red line at zero corresponds to perfect calibration (i.e., LR(LR) = LR).  The blue line is the fiducial median log discrepancy.  The black and cyan lines are upper and lower .95 pointwise and simultaneous fiducial confidence intervals, respectively, for the log discrepancy.  For this `real NFI casework data' simulation, $m_{u} = 2$, $m = 3$, $n = 659$, and $m_{i} = 3$.}\label{calibration_NFI_data}
\vspace{-.15in}
\end{figure}

Once again, the resulting sampling distributions of the methods are presented as box plots in Figure \ref{boxplots_NFI_data}.  The fiducial distributions of the AUC to assess discrimination effectiveness between the hypotheses are presented in Figure \ref{auc_NFI_data}, and the calibration analysis is displayed in Figure \ref{calibration_NFI_data}.  What is most noteworthy about the results of this simulation design is that they are largely unchanged from those of the synthetic simulation design (with matching sample sizes).  This suggests that the assumed data generating models are reasonable approximations to this real casework data, with respect to quantifying the evidence in favor of the competing hypotheses, $H_{d}$ and $H_{p}$.  Likewise, the LR and BF exhibit the same deficiencies that they did with the synthetic data.  However, the GFF tends to less extreme values than it did for the synthetic data, most noticeably for the $H_{d}$ true scenario.  

The less extreme GFF values may indicate that the GFF is less robust to model misspecification than the BF or LR.  Or it may indicate that a Gaussian random effect for the alternative source model is a better approximation for describing the NFI casework data (since the BF and LR were constructed on this assumption).  Nonetheless, the calibration of the GFF values presented in Figure \ref{calibration_NFI_data} suggests that the GFF is well calibrated for values below $10^{7}$.  The degradation in calibration for values greater than $10^{7}$ seems to be driven by 3 very large values for GFF under $H_{d}$ true in the simulation of 3,000 data sets.  Under $H_{d}$ true for the real NFI casework data, the unknown source data is neither (truly) associated with the specific nor alternative sources.  

Accordingly, it is possible that 1 in a 1,000 of the unknown source data looks very different from the alternative source data (and maybe not so different from the specific source data) in which case the denominator of the GFF is very small, so that the GFF value becomes excessively large.  We also see a similar phenomenon occurring for LR and BF (depending on the prior).  This would not happen for the simulated synthetic data simulations because under $H_{d}$ the unknown source data is actually generated from the alternative source.

In forensic identification of source applications, and particularly for those that rely on such small sample sizes, it is very important that the inferential methods being used are appropriately calibrated to reflect the strength of evidence provided by the data.  Accordingly, for small sample sizes ($m = 3$ and $m_{u} = 2$, in this case) practitioners should be very skeptical of any tool that conveys extreme confidence in favor of either of the competing hypothesis.

\section{Concluding remarks}

The motivations for this research and the writing of this manuscript are multifaceted.  The use of the BF or LR in the context of forensic identification of source applications is problematic.  Given the high stakes nature of such applications in criminal justice systems around the world, the statistics community must take responsibility for both communicating the dangerous shortcomings of these methods that are in widespread use, and for developing new methods that overcome such shortcomings.

In regards to the BF, the entire notion of ``reasonableness'' has no meaning in the context of subjectivist Bayesian prior specification/choice, especially in an adversarial scenario (e.g., prosecution versus defense).  Furthermore, while we observed the BF to be effective at discriminating between $H_{d}$ and $H_{p}$, the BF values were highly influenced by the choice of prior and they were not calibrated to represent the strength of evidence they appeared to convey.  In regards to the LR, ratios of likelihood functions evaluated at MLEs computed from excessively small data sets are very unstable, the LR values fell short in their ability to discriminate between $H_{d}$ and $H_{p}$, and they were poorly calibrated.  We have provided evidence to demonstrate these assertions empirically, and on real casework data, and we have constructed and validated a GFF as an alternative methodological approach and tool that does not suffer from the demonstrated deficiencies in the BF and LR.

Lastly, there is an argument to be made that the shortcomings in the BF approach can be remedied via the construction of {\em objective} priors (however that is to be defined).  To this point, in reference to equation (\ref{gfd}), the GFF can be interpreted precisely as a BF arising from a particular choice of {\em objective}, data-driven priors.

\section{Appendix}

In this section, the details of the BF and LR specification and computations are given.  These details for the BF are as in \cite{ommen2017}, \cite{ommen2019}.  Assuming the posterior distributions of $\theta_{s}$ and $\theta_{a}$ are independent, the BF from equation (\ref{BF}) is expressed as
{\footnotesize
\begin{equation}\label{BF_compute}
\begin{split}
\text{BF} & = \frac{\int\int f_{s}\big(\{y_{u,j}\} \mid \theta_{s}\big) \cdot \pi_{s}\big(\theta_{s} \mid \{y_{s,k}\}\big) \cdot \pi_{a}\big(\theta_{a} \mid \{y_{a,i,k}\}\big) \ d\theta_{s} \ d\theta_{a}}{\int\int f_{a}\big(\{y_{u,j}\} \mid \theta_{a}\big) \cdot \pi_{s}\big(\theta_{s} \mid \{y_{s,k}\}\big) \cdot \pi_{a}\big(\theta_{a} \mid \{y_{a,i,k}\}\big) \ d\theta_{s} \ d\theta_{a}} \\
%& = \frac{\int\int \frac{ f_{s}\big(\{y_{u,j}\} \mid \theta_{s}\big) }{ f_{a}\big(\{y_{u,j}\} \mid \theta_{a}\big) } \cdot f_{a}\big(\{y_{u,j}\} \mid \theta_{a}\big) \cdot \pi_{s}\big(\theta_{s} \mid \{y_{s,k}\}\big) \cdot \pi_{a}\big(\theta_{a} \mid \{y_{a,i,k}\}\big) \ d\theta_{s} \ d\theta_{a}}{\int\int f_{a}\big(\{y_{u,j}\} \mid \theta_{a}\big) \cdot \pi_{s}\big(\theta_{s} \mid \{y_{s,k}\}\big) \cdot \pi_{a}\big(\theta_{a} \mid \{y_{a,i,k}\}\big) \ d\theta_{s} \ d\theta_{a}} \\
%& = \int\int \frac{ f_{s}\big(\{y_{u,j}\} \mid \theta_{s}\big) }{ f_{a}\big(\{y_{u,j}\} \mid \theta_{a}\big) } \cdot \frac{ f_{a}\big(\{y_{u,j}\} \mid \theta_{a}\big) \cdot \pi_{s}\big(\theta_{s} \mid \{y_{s,k}\}\big) \cdot \pi_{a}\big(\theta_{a} \mid \{y_{a,i,k}\}\big) }{\int\int f_{a}\big(\{y_{u,j}\} \mid \theta_{a}\big) \cdot \pi_{s}\big(\theta_{s} \mid \{y_{s,k}\}\big) \cdot \pi_{a}\big(\theta_{a} \mid \{y_{a,i,k}\}\big) \ d\theta_{s} \ d\theta_{a}} \ d\theta_{s} \ d\theta_{a} \\
& = \int\int \frac{ f_{s}\big(\{y_{u,j}\} \mid \theta_{s}\big) }{ f_{a}\big(\{y_{u,j}\} \mid \theta_{a}\big) } \cdot \pi_{d}\big(\theta_{s},\theta_{a} \mid \{y_{s,k}\}, \{y_{a,i,k}, y_{u,j}\} \big) \ d\theta_{s} \ d\theta_{a}, \\
\end{split}
\end{equation}
}where
{\footnotesize
\[
\pi_{d}\big(\theta_{s},\theta_{a} \mid \{y_{s,k}\}, \{y_{a,i,k}, y_{u,j}\} \big) := \frac{ f_{a}\big(\{y_{u,j}\} \mid \theta_{a}\big) \cdot \pi_{s}\big(\theta_{s} \mid \{y_{s,k}\}\big) \cdot \pi_{a}\big(\theta_{a} \mid \{y_{a,i,k}\}\big) }{\int\int f_{a}\big(\{y_{u,j}\} \mid \theta_{a}\big) \cdot \pi_{s}\big(\theta_{s} \mid \{y_{s,k}\}\big) \cdot \pi_{a}\big(\theta_{a} \mid \{y_{a,i,k}\}\big) \ d\theta_{s} \ d\theta_{a}}
\]
}is the posterior distribution of $(\theta_{s}, \theta_{a})$ under the defense hypothesis that the unknown source data are generated from the alternative.  Note that this is simply a method for computing the BF, and it does not favor one hypothesis over another.

The random effects term in (\ref{dge_GF_a}) is assumed to follow a multivariate Gaussian distribution in \cite{ommen2017}, and they construct the following conjugate priors for the various parameters.
{\footnotesize
\begin{equation}\label{priors}
\begin{split}
\mu_{s} & \sim \text{N}_{p}(\mu_{\pi},\Sigma_{b}) \\
AA' & \sim \text{inv-Wishart}_{p}(\Sigma_{e},\nu_{e}) \\
\mu_{a} & \sim \text{N}_{p}(\mu_{\pi},k\Sigma_{b}) \\
BB' & \sim \text{inv-Wishart}_{p}(\Sigma_{b},\nu_{b}) \\
CC' & \sim \text{inv-Wishart}_{p}(\Sigma_{e},\nu_{e}), \\
\end{split}
\end{equation}
}where $k$ is some scalar.  Particularly with small samples sizes for the observed specific and unknown source data, even small variations in the data can lead to numerically unreliable BF values, especially due to the light tails of the Gaussian likelihood function.  Accordingly, from these priors it follows that it is most consistent with a belief in the prosecution hypothesis to set as diffuse as possible the specific source priors so that the unknown source data is as consistent as possible with the specific source posterior distribution.  This is done by choosing large components for $\Sigma_{b}$ for the prior on $\mu_{s}$ and small degrees of freedom parameter $\nu_{e}$ for the prior on $AA'$.  Conversely, it is most consistent with a belief in the defense hypothesis to choose small components for $\Sigma_{b}$ and a large $\nu_{e}$ so as to make the unknown source data appear as distinct as possible from the specific source posterior distribution.  In the simulation studies that follow, we construct priors from the extremes of both hypotheses in order to illustrate the excessive range in variation of the resulting BF values.  

Recall from the computational expression of the BF in (\ref{BF_compute}), the unknown source data is appended to the alternative source data.  With the updated $\{y_{a,i,k}\} = \{y_{a,i,k}, y_{u,j}\}$ and denoting $m_{n+1} := m_{u}$, the conditional posteriors resulting from the priors in (\ref{priors}) are
{\footnotesize
\begin{equation}\label{gibbs_alg}
\begin{split}
\mu_{s} \mid \{y_{s,k}\}, AA' & \sim \text{N}_{p}\big( M^{-1}L, M^{-1}\big) \\
AA' \mid \{y_{s,k}\}, \mu_{s} & \sim \text{inv-Wishart}_{p}\big( S_{s} + \Sigma_{e}, \nu_{e} + m \big) \\
\mu_{a} \mid \{y_{a,i,k}\}, BB', CC' & \sim \text{N}_{p}\big( Q^{-1}R, Q^{-1}\big) \\
CV_{i,k} \mid CC' & \sim \text{N}_{p}( 0, CC') \\
BB' \mid \{y_{a,i,k}\}, \{CV_{i,k}\}, \mu_{a} & \sim \text{inv-Wishart}_{p}\big( S_{v} + \Sigma_{b}, N+m_{n+1} + \nu_{b}\big) \\
BT_{i} \mid BB' & \sim \text{N}_{p}( 0, BB') \\
CC' \mid \{y_{a,i,k}\}, \{BT_{i}\}, \mu_{a} & \sim \text{inv-Wishart}_{p}\big( S_{a} + \Sigma_{e}, N+m_{n+1} + \nu_{e}\big), \\
\end{split}
\end{equation}
}where $S_{s}$ is defined in (\ref{S_s}), $S_{a}$ is defined in (\ref{S_a}) with an additional $m_{n+1}$ terms corresponding to the $\{y_{u,j}\}$ components, and
{\footnotesize
\[
\begin{split}
M & := m(AA')^{-1} + \Sigma_{b}^{-1} \\
L & := m(AA')^{-1}\bar{y}_{s,\cdot} + \Sigma_{b}^{-1}\mu_{\pi} \\
Q & := (N+m_{n+1})\big(BB' + CC'\big)^{-1} + (k\Sigma_{b})^{-1} \\
R & := (N+m_{n+1})\big(BB' + CC'\big)^{-1}\bar{y}_{a,\cdot,\cdot} + (k\Sigma_{b})^{-1}\mu_{\pi} \\
S_{v} & := \sum_{i=1}^{n+1}\sum_{k=1}^{m_{i}} (y_{a,i,k} - \mu_{a} - CV_{i,k}) (y_{a,i,k} - \mu_{a} - CV_{i,k})'.
\end{split}
\]
}To compute the joint posterior distribution of all the model parameters, we wrote a custom Gibbs sampler that iterates according to the updates enumerated in (\ref{gibbs_alg}).  This code is available at \verb1https://jonathanpw.github.io/research.html1.

The LR is constructed from Chapter 7.2 of \cite{ommen2017_dissertation} as,
{\footnotesize
\begin{equation}\label{LR_computation}
LR := \frac{ f_{s}\big(\{y_{u,j}\} \mid \widehat{\theta}_{s}^{\star}\big) \cdot f_{s}\big(\{y_{s,k}\} \mid \widehat{\theta}_{s}^{\star}\big) \cdot f_{a}\big(\{y_{a,i,k}\} \mid \widehat{\theta}_{a}\big) }{ f_{a}\big(\{y_{u,j}\} \mid \widehat{\theta}_{a}^{\star}\big) \cdot f_{s}\big(\{y_{s,k}\} \mid \widehat{\theta}_{s}\big) \cdot f_{a}\big(\{y_{a,i,k}\} \mid \widehat{\theta}_{a}^{\star}\big) },
\end{equation}
}where $\widehat{\theta}_{s}^{\star}$ is the MLE of the specific source parameters $\theta_{s} = \{\mu_{s},AA'\}$ from the pooled data $\{y_{s,k}, y_{u,j}\}$ based on the the prosecution hypothesis, $\widehat{\theta}_{s}$ is the MLE of $\theta_{s}$ from the data $\{y_{s,k}\}$, $\widehat{\theta}_{a}^{\star}$ is the MLE of the alternative source parameters $\theta_{a} = \{\mu_{a},BB',CC'\}$ from the pooled data $\{y_{a,i,k}, y_{u,j}\}$ based on the the defense hypothesis, and $\widehat{\theta}_{a}$ is the MLE of $\theta_{a}$ from the data $\{y_{a,i,k}\}$.  The \verb1lme1 function from the \verb2nlme2 R package \citep{pinheiro2019} is used to compute the MLE for each of the parameters.

{\spacingset{.5}\footnotesize
\bibliographystyle{agsm}
\bibliography{references}
}

\end{document}